\documentclass[aps,twocolumn,preprintnumbers,amsmath,amssymb]{revtex4}

\newcommand \be  {\begin{equation}}
\newcommand \beq {\begin{equation}}
\newcommand \bea {\begin{eqnarray} }
\newcommand \ee  {\end{equation}}
\newcommand \eeq {\end{equation}}
\newcommand \eea {\end{eqnarray}}
\newcommand{\beqa}{\begin{eqnarray}}
\newcommand{\eeqa}{\end{eqnarray}}

\newcommand{\nn}{\nonumber\\}

\usepackage{amsmath, amsfonts, amssymb,graphicx,bm}

\begin{document}                % End of preamble and beginning of text

\title{ Statistical-mechanical formulation of Lyapunov exponents}  
\author{Sorin T\u{a}nase-Nicola and Jorge Kurchan}
\affiliation{ 
 P.M.M.H. Ecole Sup{\'e}rieure de Physique et Chimie Industrielles,
\\
10, rue Vauquelin, 75231 Paris CEDEX 05,  France}
\date{\today}

\begin{abstract}
We show how the Lyapunov exponents of a dynamic system 
can in general be expressed in terms of 
the free energy  of a (non-Hermitian) quantum many-body problem.
This puts their study  as a problem  of statistical
mechanics, whose intuitive concepts and  techniques of approximation
 can hence  be borrowed.
\end{abstract}
%\pacs{75.10.Nr, 64.60.Cn, 64.70.Pf, 11.17.+y}
\maketitle

%\tableofcontents

\section{Introduction}
\label{Introduction}
% \vspace{.5cm}

Lyapunov exponents are an important tool for the characterization
of dynamical systems. 
Their very definition has a strong statistical mechanical
flavor, as it involves `extensivity in time' of  certain quantities 
regardless of the initial
(`border') conditions.  It is then natural to seek to express them
in an explicitly statistical mechanical way, in terms of a partition function.
 
In this paper, building upon previous work, mainly  of Graham
\cite{Gozzi94} 
and Gozzi \cite{Graham88},
we show how the study of Lyapunov exponents, as well as Ruelle's angle,
can be cast as a quantum many-body problem --- in fact
a
 rather standard one,
except  that the Hamiltonian is in general non-hermitian.
As we shall see, this does  not bring a miraculous solution to all 
calculational problems,
but is serves two purposes:
\begin{itemize}
\item
Because the problem is formulated as a standard quantum many-body one,
all
 the tools 
 developed in that wider context are available. 
Some, but not all of them have already been used 
as
 approximation
schemes for Lyapunov exponents, but others 
such as Hartree-Fock, dynamic mean-field theory, and the
renormalization group 
seem  promising.
\item
More important, general theoretical results are re-expressed  in a language
that is
 often familiar.
A typical example is when one asks whether finite-dimensional systems
have a
 limit
Lyapunov  density function $\rho(\lambda)$: In this setting 
the question becomes whether a quantum finite-dimensional system
has  an extensive free energy  for all chemical
 potentials.
Although this does not in itself prove the existence of
$\rho(\lambda)$ in the
 thermodynamic 
limit, it renders it intuitive and acceptable, at least up to the
level of
 rigor of theoretical physics.
\end{itemize}

Consider a general dynamical system:
\begin{equation}
\dot x_i=f_i({\bf x},{\bf{\eta}})  \;\;\; ; \;\;\; i=1,...,N
\label{ls0}
\end{equation}
and, in particular, the version with additive noise:
\begin{equation}
\dot x_i=f_i({\bf x}) +\eta_i 
\label{ls}
\end{equation}
where $\eta_i$ is a   Gaussian white variable with variance $2T_i$.
In the limit
 of zero noise we have  a standard dynamical system and 
for a particular form of  $f$  a 
 Hamiltonian system.
Clearly, an alternative way to study (\ref{ls}) is to go to the
 Fokker-Planck (or `Kramers', or
`Liouville', depending on the context)
description of evolution of probabilities:
\begin{equation}
\dot P({\bf x},t)=- H_{FP} P({\bf x},t)
\label{ffpp}
\end{equation}
where 
\begin{equation}
H_{FP} \equiv - \frac{\partial }{\partial x_i}\left(T
  \frac{\partial }
{\partial x_i}+f_i\right)
\label{ffpp1}
\end{equation}
Here and in what follows summation of repeated indices
is assumed, unless otherwise stated. 
The Fokker-Planck operator acts on the space of functions of the
coordinates 
${\bf x}$: it
resembles a Schr\"{o}dinger operator, although it 
is in general non-Hermitian. The noiseless limit  is subtle, and is
the subject of ergodic theory.

Introduce now two sets of fermion  $a^\dag_i$, $b^\dag_i$  and  
boson $\alpha^\dag_i$, $\beta^\dag_i$ creation operators
($i=1,...,N$); 
and the corresponding
vacuum, defined by:
\begin{equation}
a_i|-\rangle  =b_i|-\rangle =\alpha_i|-\rangle =\beta_i|-\rangle =0 \;\; ;
\;\; \forall i
\label{vac}
\end{equation}
It will turn out that all the information we search is obtained
directly from the following generalization of the Fokker-Planck
operator:
\begin{equation}
H_S=H_{FP}+ V_{kl}({\bf{x}})
(a^\dag_k a_l  + b^\dag_k b_l +\alpha^\dag_k \alpha_l  + \beta^\dag_k \beta_l )
\label{diez}
\end{equation}
where we denote:
\begin{equation}
V_{kl}({\bf{x}}) \equiv \frac{\partial f_k}{\partial x_l} 
\end{equation}
$H_S$ acts  on the product space of functions of the coordinates and 
 of the number of fermions and bosons of each type.
(Clearly, $H_S$ coincides with $H_{FP}$ when restricted to
 the
 zero fermion and boson subspace.)
{\em This is the many-body system mentioned above.}

If the $x_i$ are lattice variables, and the $f_i$ are short-range
 interactions,
then the system (\ref{diez}) defines  a quantum (non hermitian) 
theory, also having short-range interactions.
If  the system is instead off-lattice,  the $x_i$ 
 describe the position of 
the particles and the  $(a_i,b_i,\alpha_i,\beta_i)$ play
 the role of `spin' degrees of freedom
carried by the quantum particles:
if the $f_i$ are short-ranged, then  both the direct and the 
 spin-spin interaction are also short range.

An alternative strategy, that leads to the generalized Lyapunov
 exponents, is
 based on
 replicas of fermions $a^{\gamma \dag}_i$, $b^{\gamma \dag}_i$ with 
$\gamma=1,...,q$ and the
operator:
 \begin{equation}
H_q=H_{FP}+V_{kl}({\bf{x}})
(a^{\gamma \dag}_k a^\gamma_l  + b^{\gamma \dag}_k b^\gamma_l )
\label{diez1}
\end{equation}

The Lyapunov exponents are defined as follows.
The separation of two infinitesimally close trajectories (evolving
{\em under the same noise}, see Arnold \cite{Arnold86})
${\bf{x(t)}}$ and ${\bf{x(t)+y(t)}}$ is described by
the evolution of vectors in the tangent space:
\begin{equation}
y_k(t) = U_{ki}({\bf x_o},t) y_i(0)
\label{te}
\end{equation}    
 where
 ${\bf x_o}$ is the initial
condition and $U$ is defined as  the solution of the linear equation:
\begin{equation}
\dot{U}_{ki}(t) =  V_{kj}(x,t) U_{ji}(t)
\label{ue}
\end{equation}
The $N$ Lyapunov exponents $\lambda_i$  measure the rate of growth of the volume
in the tangent space. We construct
\begin{equation}
A(t) \equiv  
  U(t)U^\dag(t)
\label{AA}
\end{equation}
 and consider its eigenvalues ${\cal{A}}_1(t) \geq {\cal{A}}_2(t) ...\geq 
{\cal{A}}_N(t)$ (a set per initial condition and/or noise realization).
The Lyapunov exponents are
\begin{equation}
\lambda_i \equiv 
\lim_{t \rightarrow \infty} \frac{1}{t} \langle  \log {\cal{A}}_i \rangle 
\end{equation} 
Throughout  this paper averages $\langle  \bullet \rangle $ 
are over the noise and/or the initial condition. 
Note that the $\lambda_i$ are automatically labeled in a decreasing order
$\lambda_1\geq\lambda_2\geq...\geq\lambda_N$. 
 Their existence has been proved under
 very
 general 
conditions by Oseledec \cite{Oseledec68}.

One can in fact 
consider the generalized Lyapunov exponents $\lambda_i^q$ (GLE), of 
 interest as a 
 measure of the intermittency (see Benzi et al. \cite{Benzi85}) defined
for example as:
\begin{equation}
\lambda_i^{2q} \equiv \lim_{t \rightarrow \infty} \frac{1}{t} \log \langle  
  {\cal{A}}_i^q \rangle 
\end{equation} 
(a more standard definition will be given below in (\ref{lr})). 

Adopting the  nomenclature of disordered systems, 
we shall call the true  Lyapunov exponents {\em `quenched'}, 
 and the $q=1,2,...$ set the {\em `annealed'}
exponents. 
The quenched exponents can be formally obtained as the analytic
continuation  $q
\rightarrow 0$ (the replica trick), but we shall also
consider a direct evaluation here.

 Years ago, Graham
  \cite{Graham88} constructed a representation of the {\em annealed}
Lyapunov
  exponents using fermions valid for systems with one degree of
  freedom. 
Later, Gozzi \cite{Gozzi94} derived
a  generalization to several degrees of freedom. His approach differs
  from ours in two respects: first, his expression, if averaged over
initial conditions or noise, yields 
the 
  annealed quantities. Second, and also important, he considers the
  eigenvalues of $U$ rather than $UU^\dag$, a different (though
  potentially
interesting) quantity.
In the present treatment we derive an expression for the quenched exponents
 directly by introducing auxiliary fermions {\em
and bosons}: the technique of supersymmetry \cite{Efetov83,Efetov97}). We also
give expressions for the generalized exponents for even $q$. 
In all cases we consider the operator $UU^\dag$, rather than $U$: this
  is the reason why we need to introduce the fermions and bosons
in pairs $(a_i,b_i)$ and $(\alpha_i,\beta_i)$ (cfr. (\ref{diez},\ref{diez1})).

Lyapunov exponents appear naturally 
in the context of localization problems for quantum disordered 
 systems:   one considers exponents associated with
 the product of transfer matrices, which play  the role of the evolution 
operators.
(see e.g. \cite{Derrida87,Weigt96})
 Thus, the Green-function formalism has been 
treated with supersymmetric (SUSY) techniques  to extract 
correlations (see Balents and Fisher \cite{Balents97}, 
where there is also a detailed analysis
of the supersymmetry group and representation, and 
 Marko\v{s} \cite{Markos89} for
 a path-integral formalism). In these works 
 supersymmetry
 was used in order to obtain the Green
 function, and then  the quenched quantities were obtained
 via the  replica trick.

This paper is organized as follows:
In section \ref{formalism} we present the formalism for the expression
of usual  and generalized Lyapunov exponents. For Hamiltonian systems we 
show that the Lyapunov pairing rule follows 
trivially from a particle-hole symmetry, and we also discuss the computation 
of Ruelle's angle.
In section \ref{properties} we discuss the statistical mechanics of 
Lyapunov exponents of macroscopic systems, in particular  the existence of
a Lyapunov density function.
We then present two examples of application: in \ref{RM} 
to  problems
 of
 Random Matrices, and 
in section \ref{HMF}  to a Hamiltonian Mean Field model 
(HMF).

\section{The formalism}
\label{formalism}
\subsection{Basic quantities}

The average expansion of 
 $p$-dimensional volumes evolving with (\ref{ls}) 
in the space of phases  can be expressed,
in terms  of $U$ as:
\begin{widetext}
\begin{equation}
R_p(x_o,t)=\sum_{i_1,...,i_p} 
det \left[(\xi_{i_1}^\dag,.....,\xi_{i_p}^\dag)^\dag\left[
U^\dag U \right] (x_o,t) (\xi_{i_1},.....,\xi_{i_p})\right]
\label{nv} 
\end{equation}
\end{widetext}
where $\xi_{i_1}...\xi_{i_p}$ 
are all sets of $p$ orthonormal vectors. 
Introducing the cumulative Lyapunov exponents $\Lambda_i$ as
\begin{equation}
\Lambda_i \equiv \lambda_1+...+\lambda_i
\label{defn}
\end{equation}
one has 
\begin{equation}
\Lambda_p =\lim_{t\rightarrow \infty}\frac{1}{2t}\langle
\ln R_p(x_o,t)\rangle 
\label{que}
\end{equation}

  The expression (\ref{que}) is not suited
to be
 written as an integral of an exponential (a partition function) 
 because of the logarithm. This  is the usual quenched
versus
 annealed problem in disordered systems:
 we are interested in the average of the
 logarithm
 and not the logarithm of the average. Below
we shall overcome it by means of
supersymmetry, but we shall first also give expressions for the annealed quantities.

The generalized Lyapunov exponent (GLE)
 (see  \cite{Benzi85}) can be used to measure 
 intermittency (i.e. rare trajectories having unusual Lyapunov exponents)
and are relevant quantities  in the so-called ``thermodynamic formalism'' of 
 chaotic 
systems. They are defined via the
moments of the  $R_p(x_o,t)$:
\begin{equation}
\Lambda_p^{2q} \equiv \lambda_1^{2q}+...+\lambda_p^{2q}=
\lim_{t \rightarrow \infty}\frac{1}{t}\ln(\langle R_p(x_o,t)^q\rangle )
\label{lr}
\end{equation}
To the extent that  the moments $\langle R_p(x_o,t)^q\rangle $
are  sufficient  to reconstruct the
 distribution
 law of  $R_p(x_o,t)$ and the averages for all real $q$, 
one can  use them to find the quenched
 average 
(\ref{que}). 
In such cases the quenched quantities can be extracted from (\ref{lr})
 using 
the replica 
trick:
\begin{equation}
\Lambda_p=\frac{d \Lambda_p^{q}}{dq}(q=0)
\label{rtp}
\end{equation}
\subsection{Generalized exponents}

Let us first obtain an expression for the annealed exponents $\lambda_i^{2q}$
for integer $q$. This calculation is the closest to the construction of 
Gozzi \cite{Gozzi94}, with the important difference that we consider
the limit eigenvalues of the matrix $A=UU^\dag$ and not of $U$ itself.

 Introducing $p$
 pairs of fermions $a_i$ and $b_i$ one can 
 write:
\begin{equation}
R_p(x_o,t)=\langle -|\psi_p{\cal {T}}\left[e^{\int_0^t H_1(x(t'))dt'}\right]
\psi_p^\dag|-\rangle 
\label{nvf}
\end{equation}
where ${\cal {T}}$ denotes time-order
and
\bea
H_1(t) &\equiv & V({\bf{x}})_{ij}(a_i^{\dag} a_j + b_i^{
\dag} b_j ) \\
\psi_p &\equiv& \frac{1}{p!}\sum_{i_1,...,i_p} a_{i_1} ...a_{i_p}  b_{i_1} ... 
b_{i_p}
\label{ham22}
\eea
The explicit time-dependence of 
$H_1$,  is given  by the evolution of ${\bf x}$  Eq. (\ref{ls})
via  $V_{ij}({\bf x})$. 
More generally, one  can write, in terms of $q$ replicas of the set of 
fermions $a_i^\gamma$ and $b_i^\gamma$:
\begin{equation}
R_p^{q}(x_o,t)=\langle -|\psi_p^q {\cal {T}}\left[e^{\int_0^t H_1^q(x(t'))dt'}\right]
\psi_p^{q \dag}|-\rangle 
\label{nvf1}
\end{equation} 
where 
\bea
H^q_1(t) &\equiv & V({\bf{x}})_{ij}(a_i^{\gamma \dag} a^\gamma_j + b_i^{\gamma
\dag} b^\gamma_j ) \\
\psi_p^q&\equiv&\Pi_{\gamma=1}^{q} 
\sum_{i_1,...,i_p} a^\gamma_{i_1} ...a^\gamma_{i_p}  b^\gamma_{i_1} ... 
b^\gamma_{i_p}
\label{ham1}
\eea
(see appendix \ref{A}). This is not the final expression, since we have not
yet imposed that ${\bf{x}}$ evolves according to the equation of motion.

The use of two sets ($a$'s and $b$'s) of fermions
 are  needed in order to follow the rate of growth of the 
norm of vectors in the tangent space and not the evolution of the
 vectors themselves: we are interested in the
 eigenvalues 
of $U^\dag U$ and 
not in those of $U$. (This is obviously not a problem in dimension one
\cite{Graham88}).
In dimensions larger than one,  Gozzi \cite{Gozzi94} 
 introduced one family of fermions and thus
studied  the eigenvalues of $U$:  these
do not in 
general coincide with the Lyapunov exponents
 (see references \cite{Orszag87} and \cite{Crisanti93}
 for a discussion), although they surely give relevant information.
In section \ref{RM} we will  explicitly show 
in an example how the $a$ and $b$-fermions
interfere in a non-trivial way.

We now write  an expression for the
average over noise and/or initial conditions of 
 (\ref{nvf1}) using the information 
that ${\bf x(x_o,t)}$ evolves according to  
 (\ref{ls}), and the probability density follows (\ref{ffpp1}).
One way to do this is to  express the weight of each trajectory 
as: 
\begin{widetext}
\begin{equation}
\langle R_p^{q}({\bf x_o},t) 
\rangle=\left\langle \int \; 
 Dx \;  \delta(x-{\bf x(x_o,\eta,}t')) \; 
\langle -|\psi_p^q \left[e^{\int_0^t H_1^q(x(t'))dt'}\right]
\psi_p^{q \dag}|-\rangle \right\rangle_\eta 
\end{equation}
where $Dx$ means the flat functional integral over trajectories and the 
delta-function imposes that $x(t)$ satisfies the equation of motion.
This passage is just the standard textbook
 exercise of going from a Langevin to a Fokker-Planck description
(see for example \cite{zinn-justin96}). The result is that
the probability evolves through the Fokker-Planck equation, and we have:
\begin{equation}
\langle R_p^{q}({\bf x_o},t) \rangle =\langle 1|\otimes \langle -|\psi_p^q
  \left[e^{-t H_R} \right] |\psi_p^{q\dag}
|-\rangle  \otimes |{\bf x_o}\rangle 
\label{fp}
\end{equation}
\end{widetext}
where $H_q=H_{FP}-H_1^q$ is given in (\ref{diez1}).   
The ket  $\langle 1|$ is the flat measure $\langle 1|{\bf x}\rangle=1$
and $\langle {\bf x}|{\bf x_o}\rangle =\delta({\bf x}-{\bf x_o})$
is the distribution associated to ${\bf x_o}$.
Note that the time-ordering is automatic, as the evolution of ${\bf x}$ is
taken care of by $H_{FP}$.
In the limit 
 $t \rightarrow \infty$ the logarithm of $\langle R_p^{q}({\bf x_o},t)
\rangle$
is given by the smallest eigenvalue of $H_q$ in the subspace
not orthogonal to the vectors  $\langle 1|\otimes \langle -|\psi_p^q$ and 
$\psi_p^{q\dag}|-\rangle  \otimes |{\bf x_o}\rangle $.

We can now describe a practical algorithm for computing the GLE's. 
First we have to identify the smallest  invariant subspace 
containing the states appearing to the right and to the left of (\ref{fp}).
Clearly,  the fermion numbers 
$N_a^\gamma = \sum_i a_i^{\gamma \dag} a_i^{\gamma}$
and $N_b^\gamma = \sum_i b_i^{\gamma \dag} b_i^{\gamma}$ are conserved,
and the calculation of $\langle R_p^{q}({\bf x_o},t)\rangle$
 involves working
in the subspace $N_b^\gamma=N_a^\gamma=p$.
 Furthermore, because  $H_q$ and $\psi_p^{q}$ commute with the operators 
${\cal {P}}^\gamma$ defined by:
\begin{equation}
 {\cal {P}}^\gamma a_i^\gamma {\cal {P}}^{\gamma \dag} = -b_i^\gamma \;\;\; ; \;\;\; 
 {\cal {P}}^\gamma b_i^\gamma {\cal {P}}^{\gamma \dag} = a_i^\gamma
\label{excha}
\end{equation}
we should look into the subspace of eigenfunctions symmetric under
exchange of $a$'s and $b$'s, 
having eigenvalue one under the
${\cal {P}}^\gamma$: {\em here is where
  two families of fermions get mixed}.
 Because the  quantity (\ref{fp}) is by construction positive, one can
see  that the eigenvalue having
 the smallest real part within this subspace  has zero 
imaginary part.
\subsubsection{Variational calculations}
  A variational approach can immediately be implemented  
for the calculation of the generalized (and, with an act of
 faith, the usual) Lyapunov exponents.
Since one is looking for the lowest eigenvalue of $H_q$ within a given 
subspace, one can use a variational trial function  $|\varphi \rangle$
\begin{equation}
\Lambda^{2q}_i \sim \min_{\varphi} {\mbox{Re}}  \frac{
\langle \varphi |H_q|\varphi \rangle }{\langle \varphi |\varphi \rangle}
\end{equation}
If the family of variational functions is parametrized for every $i,q$,
one obtains an approximation which is an explicit function of $q$:
one can then envisage computing also an approximation for the
 usual Lyapunov exponents via  (\ref{lr}).

\subsection{Ordinary (quenched) Lyapunov exponents}
\label{susy}

In this section we construct an expression for the (quenched) Lyapunov
exponents using a supersymmetry formalism.
For the purposes of clarity,
we shall do so in two steps: first a na\"{\i}ve calculation
that does not take into account the convergence of the sums, but
is closer to the standard supersymmetry treatments in other contexts.
Next, we use the Borel transform technique to work more properly.

\subsubsection{Careless calculation}

We introduce a set of fermion and boson  operators as in (\ref{vac}) and
(\ref{diez}), and define the number operators: 
\begin{eqnarray}
 N_a  =a^\dag_k a_k \;\;& ;& \;\;   N_b  =  b^\dag_k b_k     \nn
N_\alpha=\alpha^\dag_k \alpha_k  \;\;&;&\;\;  N_\beta =\beta^\dag_k
 \beta_k \nn 
 N_{bos}=\frac{N_\alpha+N_\beta}{2}& ;&  N_{fer}=\frac{N_a+N_b}{2}\nn
\bar{N}=&N_{fer}&+N_{bos}
\end{eqnarray}
which will commute with all other operators.
We shall also need:
\begin{equation}
  f \equiv
 a^\dag_k b^\dag_k \;\;\; ; \;\;\; \bar f \equiv \alpha^\dag_k \beta^\dag_k
\label{mann}
\end{equation}
Let us now introduce, for a given trajectory ${\bf x(t)}$, 
the quantity:
\begin{equation}
H_1^S(t)= V_{kl}({\bf{x}})
(a^\dag_k a_l  + b^\dag_k b_l +\alpha^\dag_k \alpha_l  + \beta^\dag_k \beta_l )
\end{equation}
With these notations we compute  $Z(\mu,{\bar\mu})$ as:
\begin{equation}
\langle -| e^{f^\dag+ {\bar f}^\dag} \left( {\cal {T}} e^{t (H_1^S-{\bar \mu}
    N_{bos}-\mu 
N_{fer})} 
\right) e^{f+  {\bar f}} (-1)^{N_{Bos}} |-\rangle  
\label{fon}
\end{equation}
an one can easily show that (Appendix \ref{bosons}):
\begin{equation}
 Z(\mu,{\bar\mu})= \frac{ det [ 1+e^{-\mu t} A]}{ det [ 1+e^{-{\bar\mu} t}A]}
\label{ufa}
\end{equation}
where $A(t)$ is associated with the trajectory ${\bf x}$. 
This function will generate all the Lyapunov exponents for the trajectory as:
\begin{equation}
G(\mu) \equiv - \lim_{t \rightarrow \infty}
 \left.
 \frac{1}{t} \frac{\partial Z(\mu,{\bar\mu})}{\partial
 \mu}\right|_{{\bar\mu}=
\mu}=  \lim_{t \rightarrow \infty} \sum_{j=1}^{N}\frac{e^{-\mu t} {\cal{A}}_j}{1+e^{-\mu t} {\cal{A}}_j}
\label{liapu1}
\end{equation}
since for large $t$, $G(\mu)$ is a ladder
 with steps
 at the values of $\mu$ corresponding to the $\lambda_i$'s 
(i.e. it is the integral of the Lyapunov distribution function).
Just as in the previous subsection, we wish to calculate
 $\langle  G(\mu) \rangle $.
Again, this is directly done in the Fokker-Planck formalism by the quantity:
\begin{equation}
\langle G(\mu)\rangle  \equiv -\lim_{t \rightarrow \infty}
 \left. \frac{1}{t} \frac{\partial 
\langle Z(\mu,{\bar\mu})\rangle }{\partial \mu}\right|_{{\bar\mu}=\mu}
\label{pag}
\end{equation}
with $\langle  Z(\mu,{\bar\mu})\rangle  $ written as:
\begin{equation}
 \langle \phi_L| 
  e^{-t (H_S+\mu N_{fer}+{\bar \mu} N_{bos})}(-1)^{N_{Bos}}  |\phi_R\rangle 
\label{formula}
\end{equation}
where $H_S=H_{FP}-H_1^S$  is given in (\ref{diez}). 
The left and right eigenvectors are
\begin{eqnarray}
 |\phi_R\rangle &=&\int d{\bf x_o}\;P({\bf x_o})\;
 e^{f+  {\bar f}}|-\rangle \otimes 
|{\bf x_{o}}\rangle \nn
\langle \phi_L| &=& 
 \langle -|\otimes \langle -| e^{f^\dag+  {\bar f}^\dag} 
\end{eqnarray}
where $P({\bf x_o})$ is the initial condition distribution.
We can also write $\langle G(\mu)\rangle $ as an expectation value:
\begin{equation}
\langle G(\mu)\rangle=\langle \phi_L|
   e^{-t \{H_S+\mu \bar N\}} 
\; N_{fer}(-1)^{N_{bos}} |\phi_R\rangle 
\label{fon2}
\end{equation}

Equation (\ref{ufa}) can be expressed as a formal series in powers
of $\exp(-t\mu)$ and $\exp(-t{\bar \mu})$. This same series
is reproduced by (\ref{fon}), or, in averaged version, (\ref{formula}).
Each term of the form $\exp[-t (n_1 \mu+ n_2 {\bar \mu})]$ corresponds
to an expectation value in the subspace of $n_1$ fermions and $n_2$ bosons.
Clearly, (\ref{ufa}) has a very small convergence radius:  this is because
the number of bosons, unlike the number of fermions, is unlimited.
In other words, expressions (\ref{ufa}), (\ref{fon}) and (\ref{formula})
have only a formal meaning.
 One can still work with them if at the end of
a derivation one can resum exactly the series, in which case one has 
in fact performed the analytic continuation in $\mu$ and $\bar \mu$.

\subsubsection{Borel transform}

Let us now give a more proper construction to the supersymmetric
expression of the Lyapunov exponents. It will turn out that 
the formalism that emerges is not much more complicated than the one in the
previous paragraphs.

Briefly, the Borel transform technique consists of going from the 
formal series:
\begin{equation}
h(y) \sim a_0 + a_1 y + a_2 y^2 + ...
\end{equation}
to its convergent transform
\begin{equation}
h^B(y) = a_0 + \frac{a_1}{1!} y + \frac{a_2}{2!} y^2 + ...
\end{equation}
 which can then be inverted.
In our case, we shall take
\begin{equation}
h(y)  \sim a_0 + a_1 y + a_2 y^2 + ...=
 \frac{ det [ 1+y e^{-\mu t} A]}{ det [ 1+ y e^{-{\bar\mu} t}A]}
\label{ufaB}
\end{equation}
(cfr. Eq. (\ref{ufa})), and define as the Borel-transformed partition
function:
\begin{equation}
Z^{B}(\mu,{\bar\mu}) = \left. h^B(y)\right|_{y=1}
\end{equation}
that is, we are dividing by  $(n_1+n_2)!$ the term having $n_1$ fermions
and $n_2$ bosons.
Repeating the construction above, it is easy to see that:
 \begin{equation}
\langle Z^{B}(\mu,{\bar\mu}) \rangle=
 \langle \phi_L^B| 
  e^{-t (H_S+\mu N_{fer}+{\bar \mu} N_{bos})} (-1)^{N_{Bos}} |\phi_R^B\rangle 
\label{formulaB}
\end{equation}
which is just like (\ref{formula}) except that 
the left and right eigenvectors are
\begin{eqnarray}
 |\phi_R^B\rangle &=&\int d{\bf x_o}\;P({\bf x_o})\;
 g({f+  {\bar f}})|-\rangle \otimes 
|{\bf x_{o}}\rangle \nn
\langle \phi_L^B| &=& 
\int  \langle -|\otimes \langle -| g({f^\dag+  {\bar f}^\dag}) 
\end{eqnarray}
where the function $g(x)$ is:
\begin{equation}
g(x)=\sum_{n=0}^\infty \frac{x^n}{n!\sqrt{n!}}
\end{equation}
We can now define as before:
\begin{equation}
\langle G^B(\mu)\rangle  \equiv - \lim_{t \rightarrow \infty}
 \left. \frac{1}{t} \frac{\partial 
\langle Z^B(\mu,{\bar\mu})\rangle }{\partial \mu}\right|_{{\bar\mu}=\mu}
\label{pag1}
\end{equation}
which we can also write  as an expectation value:
\begin{equation}
\langle G^B(\mu)\rangle= \langle \phi_L^B| 
 e^{-t \{H_S+\mu \bar N\}} 
\; N_{fer} (-1)^{N_{bos}} |\phi_R^B\rangle 
\label{fon2b}
\end{equation}
Interestingly enough, we can retrieve the information {\em directly}
(without the need to antitransform)
from $G^B(\mu)$, since one can show (Appendix \ref{borsum}) that: 
\begin{equation}
G^B(\mu)=\sum_{j=1}^{N}\theta(\lambda_j-\mu)
\label{bonexp}
\end{equation}
again gives the same ladder function.
\subsubsection{Symmetries}
Although we shall not make here much use of the symmetries in the problem,
let us briefly mention them for completeness.
Clearly,   the Hamiltonian $H_{S}$ 
is  invariant under
all the  (supersymmetric) transformations rotating
simultaneously the  $\alpha_i,\beta_i,a_i,b_i$
so as to leave the quadratic form 
$f+{\bar{f}}$ and the fermion-boson part of $H_S$ invariant.
Then, expectation values can be written in the standard form:
\begin{equation}
\langle O \rangle =
Tr [(-1)^{N_{fer}} {\bf C}  O]
\end{equation}
with ${\bf C}$ supersymmetric:
\begin{equation}
 {\bf C} \equiv   e^{-t \{H_S+(\mu + i\pi) (N_{fer}+ N_{bos})\}}
|\phi_R^B\rangle \langle \phi_L^B| 
\end{equation}
Both the original and 
 the Borel transformed versions have the same symmetries.

One can show that  supersymmetry is responsible for the  fact that
$Z(\mu,\mu)$ and $Z^B(\mu,\mu)$ are independent of $\mu$: the  
constancy of the 
normalization  is indeed the underlying reason why we can
 use the method to obtain
quenched averages. 
 (See Ref. \cite{Balents97}) for a discussion in detail
of the supersymmetry group  and representations).

\subsection{Discussion}

The calculation of generalized exponents $\lambda_i^{2q}$ is done,
as we have seen, by computing the lowest eigenvalue within a subspace
of the Hilbert space. This is because the large-$t$ limit automatically
projects onto the corresponding eigenstate.
Approximate and numerical methods for the estimation of the ground state
of Schr\"odinger-like operators abound in the literature, we have already 
mentioned
the variational principle.

If one wishes to extend these results for the quenched exponents,
an analytic continuation to $q \rightarrow 0$ is needed.
This is easily done 
(although with a leap of faith) when an explicit expression for all even 
$q$ is available. Such will be the case in a variational, a perturbative or a
mean-field computation.

On the other hand, the supersymmetry method yields the Lyapunov exponents
without the need of any continuation.
However, there is a price to pay: expressions (\ref{fon2}) and  (\ref{fon2b})
 involve a sum
of terms within subspaces of any number of bosons. In the Borel transformed
version (\ref{fon2b}) this sum is convergent for all finite $t$. 
However, one can see
that the largest
term corresponds to a boson number of the order of $\exp(\lambda_1 t)$,
and this number grows as we consider larger times.
In other words, we can only  perform this sum for finite $t$,
and only then make $t \rightarrow \infty$.
Again, this is no problem if an analytic expression is available (perhaps
as a result of an approximation), but it does seem problematic to attack
a problem numerically this way.
\subsection{Hamiltonian systems: Pairing rule and  Ruelle Angle}
Damped Hamiltonian systems are a particular case of the dynamics (\ref{ls}), which can be written
as
\begin{eqnarray}
{\dot q} &=& -\frac{\partial {\cal H}}{\partial p_i} \nonumber \\
 {\dot p} &=&\frac{\partial {\cal H}}{\partial q_i} - \gamma p_i + \gamma^{\frac{1}{2}} \eta_i
\end{eqnarray}
where $\gamma$ measures the intensity of the coupling to the bath.
The simplectic structure of Hamilton's equations
has consequences.

\subsubsection{Pairing of exponents}

One of the proprieties we can easily infer from our formulation is the
 pairing-rule for the Lyapunov systems. This pairing rule was proved
 for a
 class of dynamical systems (named `quasi-hamiltonian' by Dressler
 \cite{Dressler88}); in our formalism it is the result of a
 particle-hole symmetry. 
This is most easily seen for the annealed exponents. 
In the case
\begin{equation}
{\cal H} = \frac{p_i^2}{2} + {\cal V}({\bf q})
\end{equation}
the $V_{ij}$ read, 
\begin{equation}
V=\left(\begin{matrix}0&\boldsymbol{1}\cr-\frac{\partial^2 {\cal V}}{\partial
      q_i \partial q_j}
&-\gamma\boldsymbol{1}\end{matrix}\right)
\end{equation}

Denoting $x_i=q_i$ for ($i=1,...,N$) and $x_i=p_i$ for ($i=N+1,...,2N$), 
we consider the transformation:
\begin{equation}
\bar a^{\dag}_i=\epsilon_{ij}a_j \quad \bar b^{\dag}_i=\epsilon_{ij}b_j
\end{equation}
where 
\begin{equation} 
\epsilon = \left(\begin{matrix}\boldsymbol{0}&-\boldsymbol{1}
\cr\boldsymbol{1}&\boldsymbol{0}\end{matrix}\right)
\end{equation}
(here and in what follows $\boldsymbol{0}$ and $\boldsymbol{1}$ are
 the
 null matrix and the identity matrix in the $N\times N$ space).

Under this transformation the quantities from (\ref{nvf}) transform as:
\bea
{\bar {H_q}} \rightarrow H_q -2q\gamma(N-N_{fer}) \\
{\bar {\psi_p} |\bar 0\rangle}  = \psi_{N-p}|0\rangle  
\eea
This particle-hole symmetry, a consequence of the symplectic structure of
 the evolution operator, was noted by Gozzi \cite{Gozzi94}. 
One then gets:
\begin{equation}
\Lambda_p^{2q}=\Lambda_{N-p}^{2q}+(N-p)2q\gamma
\end{equation}
which in terms of the exponents becomes 
\begin{equation}
\lambda_n^{2q}=-2q\gamma-\lambda_{N-n+1}^{2q}
\end{equation}
Using (\ref{rtp}) and (\ref{defn}) we can infer that the Lyapunov exponents are symmetric around $-\frac{\gamma}{2}$.
Note that this result holds for arbitrary $\eta_i(t)$, and not only in the case in which it is 
a white noise.

\subsubsection{Ruelle's angle}

The difficulty in computing the Lyapunov exponents is due to the fact that, in the tangent 
space, any instantaneous frame generically turns. In order to define
 an angle associated with these       transformations  Ruelle
 \cite{Ruelle85}
 defined a rotation number (or Ruelle frequency -RF- after Dressler 
\cite{Dressler90}) and  proved an additive  theorem for  
it.
To construct
 this number we can use a polar decomposition of the evolution 
operator: $U=A^{\frac{1}{2}}Q$ with $A\equiv UU^\dag$   simplectic and
 symmetric and $Q$ simplectic and orthogonal:
\begin{equation}
Q=\left(\begin{matrix}X&Y\cr -Y&X\end{matrix}\right)
\end{equation}
Clearly,  $T \equiv X+iY$ is  unitary; and we can extract a
 rotation number from  $det(T)=e^{im}$  
(the determinant of an unitary matrix is a pure phase).

The Ruelle angle can be extracted directly  from $U$ (see \cite{Littlejohn86})
  in the following way 
\footnote{The argument presented by Littlejohn \cite{Littlejohn86} 
is for closed trajectories but it works the same for open ones.}: 
write the evolution operator as 
\begin{equation}
U=\left(\begin{matrix}U_A &U_B\cr U_C&U_D\end{matrix}\right)
\end{equation}
Then,  $m(t)$ will be the argument of $z(t)=det(U_A+iU_B)$. We can write:
\begin{equation}
e^{2im(t)}=
\frac{z(t)^2}{z(t)z(t)^*}=\frac{det(U_A+iU_B)^2}{det(U_AU_A^\dag+U_BU_B^\dag)}
\label{pel}
\end{equation}
where we have used the fact that, since $U$ is a real simplectic matrix,
$
U_AU_B^\dag=U_BU_A^\dag
$.

We can use fermions to write the determinant quantities as:
\begin{equation}
det(U_A+iU_B)=\langle -|
\prod_{i=1}^N
a_{q_i}\; {\cal{T}} e^{tV} \; \prod_{i=1}^N(a_{q_i}^\dag+
i a_{p_i}^\dag)|-\rangle 
\end{equation}
(here we adopt the notation $a_{q_i}$ $a_{p_i}$
 for the  fermion corresponding to 
the $i-th$  position or momentum degree of freedom  ($i= 1,..., N$).
We can further use the $b$ fermions to  obtain the square of this determinant.
To compute the denominator of (\ref{pel}), we first note that 
$O= U_AU_A^\dag+U_BU_B^\dag$ is the top-left block in 
the $A=UU^\dag$. To have his determinant in the denominator we can use the 
relation:
\begin{equation}
\frac{1}{det(1+O)}=\langle -|e^{\alpha_{q_i}\beta_{q_i}}T
 e^{tV}(-1)^{N_{bos}}e^{\alpha^{\dag}_{q_i}\beta^{\dag}_{q_i}+\alpha^{\dag}_{p_i}
\beta^{\dag}_{p_i}}|-\rangle 
\end{equation}
But, since $O$ is an $N \times N$  block of $A$, it will generically
have a projection in the direction of 
the $N$ largest eigenvalues, so for large times:
\begin{equation}
det(1+O(t))\simeq det(O(t))\simeq e^{S_{>} t}
\end{equation}
where $S_{>}$ is the sum of the largest $N$ Lyapunov exponents.
Putting everything together, we can write 
\begin{widetext}
\begin{equation}
\langle e^{2im(t)} \rangle =\langle 1|\otimes
\langle -|e^{\alpha_{q_i}\beta_{q_i}}
\prod_{k=1}^N a_{q_k} b_{q_k} \; \left[e^{tH_S}\right] \; \prod_{l=1}^N(
a_{q_l}^{\dag}
+i a_{p_l}^{\dag})(b_{q_l}^{\dag}+i b_{p_l}^{\dag})(-1)^{N_{bos}}
e^{\alpha^{\dag}_{q_i}\beta^{\dag}_{q_i}+\alpha^{\dag}_{p_i}
\beta^{\dag}_{p_i}}
|-\rangle \otimes |{\bf x_o}\rangle 
\end{equation}
\end{widetext}

Note that the quantity we have calculated is not quite the quenched 
averaged angle, but rather 
the average of the exponential.
\section{Macroscopic systems} 
\label{properties}
An active field of research is the information that  Lyapunov exponents 
can provide in  extensive systems in thermodynamic limit.
The natural object to study is the Lyapunov density function:
\begin{equation}
\rho(\mu) \equiv \sum_{i=1}^N \delta(\lambda_i-\mu)
\label{density} 
\end{equation}
or, better, the cumulative version:
\begin{equation}
C(\mu) \equiv \int_\mu^\infty 
\rho(\lambda) d \lambda
\label{CU}
\end{equation}
In particular $C(0)$ 
is the sum of positive exponents, related to the  Kolmogorov-Sinai (KS) 
entropy.
The existence of a thermodynamic limit for the Lyapunov densities 
  has been  conjectured (\cite{Ruelle82,Livi86}).
 A problem immediately arises: the Lyapunov exponents are themselves
the result of the limit $t \rightarrow \infty$, and the question as to whether
this, and the thermodynamic limits commute is not obvious.
Typical examples when they do not is when there are 
macroscopic motions that take
times that diverge with the size. We shall find a clear example 
in the  Hamiltonian mean-field model below (Section \ref{HMF}): the particles
organize in an anisotropic object which can turn collectively like a rotor:
In contact with a bath there is  diffusion of the collective angle, but, 
 since the moment of inertia
scales with $N$, the collective motion is absent if we consider 
$N \rightarrow \infty$
 {\em before} $t \rightarrow \infty$. 
  Sinai\cite{Sinai96}
 has shown that in this last order of limits the densities are
 well defined in a system of confined
 particles with pairwise interactions:
 we shall see that this result is very natural.

In the previous sections we have shown that  $C(\mu)$ is the 
large time limit of $C_t(\mu)$ with 
\begin{equation}
C_t(\mu) = G^B(\mu,t) = 
\left. \frac{1}{t} \frac{\partial \ln
 Z^B(\mu,{\bar \mu},t)}{\partial \mu}\right|_{\mu={\bar \mu}}
\label{cdef}
\end{equation}

(the logarithm in the r.h.s. has no effect, since the normalization is one).

Consider this expression: $ Z^B$ is a partition function associated
with the `quantum' hamiltonian $H_S$, 
where the time plays the role of an inverse temperature,
$\mu$ of a chemical potential and
$C_t(\mu)$ of the derivative of a free energy density with respect to the
 chemical potential
(i.e. a particle number per unit volume). The Lyapunov density is hence 
a form of compressibility.

As mentioned in the introduction, 
if the original problem is on a lattice and has nearest neighbor interactions,
the fermions and bosons are also lattice variables  interacting  with 
the  nearest
neighbor variables  (through    $V_{ij}$). We have then a   `quantum'
lattice problem with short range interactions.
On the other hand, the system could be off-lattice, and the $x_i$ be a set of
 d-dimensional vectors describing  the position of 
the particles
 interacting  via short-range 
pair forces $f_i({\bf x})=\sum_j f(x_i-x_j)$.
The variables $(a_i,b_i,\alpha_i,\beta_i)$ play the role of `spin' 
degrees of freedom
carried by quantum particles, both the
 direct and the  spin-spin interaction are also short range.

All in all, we are asking whether a quantum 
theory with short range interactions
has a good thermodynamic limit with a well defined free-energy density.
There is only one non-standard feature
if we  ask for the $t \rightarrow \infty$
 taken before the thermodynamic limit: this is, as we have seen, like asking 
in a statistical mechanical problem 
about the zero temperature limit taken before the thermodynamic limit - 
sometimes a tricky question.

The arguments on extensivity become  more subtle if we wish to study 
the thermodynamic limit of
the {\em largest} Lyapunov exponent:
this is like asking in a particle system not what is the chemical potential
 needed to create a certain particle density, but rather to create {\em a
single particle in the whole system}: clearly this is a question of order
$O(1/N)$. 
We shall return to this
 point in section \ref{HMF}.

An interesting special case is the behavior of
the Lyapunov exponents close to zero in a system with soft modes.
In the present context this concerns the properties of a statistical mechanic
problem around $\mu=0$, i.e. free from external chemical potential.
For example, from Eq. (\ref{fon2}) we have:
\begin{equation}
\langle \rho(\lambda=0) \rangle= \langle \phi_L^B| 
 e^{-t \; H_S} 
\; N_{fer} (N_{fer}+ N_{bos}) |\phi_R^B\rangle 
\label{fon22}
\end{equation}

A very intriguing possibility that immediately comes to mind when
working in
the present framework  is that of studying  universality
properties in critical points using  
renormalization-group ideas and techniques.

%\vspace{.5cm}
\section{Random Matrices}
\label{RM}
%\vspace{.5cm}

In this section we use our formalism to derive some results already 
obtained for some Random Matrix (RM) models. For brevity we shall
only do this in the pure fermion (annealed) case, although 
the supersymmetric approach can also be applied.

The main lesson we shall obtain is that these systems become, by virtue of
the disorder, interacting fermion problems.
As such, they can be very well attacked by some of the many methods
devised for such cases: Feynman diagrams of course, but also
ressumations such as Hartree-Fock.
 
\subsection{Weak disorder expansion}
\label{derrida}

The first  model where we can show the power of our approach is the 
one proposed by Derrida \& al.\cite{Derrida87}. They study the weak disorder 
expansion of the quenched Lyapunov exponents for a product of the form:
\begin{equation}
P=\prod_t U_t \quad U_t=B_o+\epsilon B
\end{equation}
where $B_o$ is a fixed matrix, $B$ is a random matrix, and 
$\epsilon$ is a small parameter.
We shall study the case with $B$ a multi-dimensional, Gaussian white noise, 
with zero mean (as the finite mean can be safely 
included in the constant matrix):
\bea
\langle B_{ij}(t)\rangle &=&0 \nn \langle B_{ij}(t_1)B_{kl}(t_2)\rangle
 &=&\sigma_{ij,kl}\delta(t_1-t_2)
\eea
Following Derrida we study the case with the matrix $B_o$ having
non-degenerate eigenvalues, well separated.

The first step in order to use our formalism is the set-up 
 a continuous-time variant
 of the problem:
\begin{equation} 
U(dt)=1+B_o dt+\epsilon B dt
\end{equation}
Next, we exponentiate this expression. Due to the non-continuous
character of the 
random term $B$ the correct form of the exponential is:
\begin{equation}
U(dt)=e^{B_o dt + B dt -\frac{1}{2} \langle B^2\rangle  dt^2}
\end{equation}
This means that the evolution of $U$ will be given by (as in Eq.
(\ref{ue})):
\begin{equation}
V_{ij}(t)=B_{oij}+B_{ij}(t)-\frac{1}{2}\sum_k\langle B_{ik}B_{kj}\rangle 
\end{equation}
We can derive now the path-integral form of $R_1^q$ as:
\begin{equation}
\langle -|\prod^{q}_{\gamma=1} a^\gamma_{i_1} ...a^\gamma_{i_n}  b^\gamma_{i_1}
... b^\gamma_{i_n}|e^{-tH} |\prod^q_{\gamma=1} a^{\gamma\dag}_{l_1}
...a^{\gamma\dag}_{l_n}
b^{\gamma\dag}_{l_1}...b^{\gamma\dag}_{l_n}|-\rangle 
\label{rd}
\end{equation}
($\gamma$ is the replica index).
We used as a Hamiltonian
\begin{equation}
H=V_{ij}(x)\sum_{\gamma=1}^q(a_i^{\gamma\dag} a_j^\gamma
    + b_i^{\gamma\dag} b^\gamma_j )
\label{hd}
\end{equation}
The average  $\langle R_1^q\rangle $ will be expressed 
by integrating the
 gaussian
  noise $B_{ij}$ in (\ref{rd}) which will transform $H$ into 
\bea
H'  &=& H_o + \epsilon^2 H_I\\
H_o&=& B_{oii}\sum_{\gamma=1}^q(a_i^{\gamma\dag} a_i^\gamma +
 b_i^{\gamma\dag} b^\gamma_i) \\
H_I &=& \frac{\langle B_{mn}B_{ij}\rangle }{2}\sum_{{\gamma'},
\gamma =1}^q(a_i^{{\gamma} \dag}
  a_m^{{\gamma'}\dag} a_n^{\gamma'} a^\gamma_j+
\nn
& & +b_i^{\gamma\dag} 
 b_m^{{\gamma'}\dag}  b_n^{\gamma'} b^\gamma_j+a_i^{\gamma\dag}
  b_m^{{\gamma'}\dag} b_n^{\gamma'} a^\gamma_j)
\label{Dham}
\eea
We have used the fact that we can diagonalize $B_o$ and work on that base 
(which will not change the fermionic states). We have also arranged
the
 creators and the destructors of fermions in  normal form. 
The expression (\ref{Dham}) 
can be proved to be the correct one by going back to the  
Suzuki-Trotter product, performing the averages there, and then reconstructing the 
continuous version.
We are now in the possession of a time-independent Hamiltonian.
Note that  after integration
of the noise we get  a non-trivial result only  because of
the presence
 of two types of fermions: without them we would have lost
 the higher
 moments of the noise.
The replica trick is easy to implement  for perturbation expansions,
since these yield
 an explicit dependence as a polynomial
 in $q$ of each term of the 
expansion. This is exactly the program we carry on now.
First, the eigenstates of $H_o$ are:
\begin{equation}
\psi_o=\prod_\gamma^q\prod_i^p a_i^\gamma b_i^\gamma
\end{equation}
and the corresponding eigenvalue is:
\begin{equation}
2q \; \sum^p_i \varepsilon_i
\end{equation}
($\varepsilon_i$ are the eigenvalues of $B_o$).
The first non-zero order in $\epsilon$ is in $\epsilon^2$;
 in order to compute
 it we must
  use  first-order perturbation theory to obtain:
\bea
\langle \psi_o|H_I|\psi^\dag_o\rangle =
-\frac{\langle B_{ij}B_{ji}\rangle }{2} (2q) + 3 q^2 
\frac{\langle B_{ii}B_{jj}\rangle }{2}
\eea

For the second order perturbation theory (which will give
 the contribution in $\epsilon^2$) we must identify the states
 connected by the perturbation. 
We can now retain the terms linear in $q$, and hence
obtain the quenched average (the coefficients of the linear term
 (see (\ref{lr})):
\bea
\varepsilon_p &=& \sum_{i=1}^p \varepsilon_i - \epsilon^2 \frac{\langle B_{ij}B_{ji}\rangle }{2}+ \nonumber \\ 
& &+\epsilon^4 \sum_{i=1}^p \sum_{l=1}^p \sum_{j=1}^p\sum_{m > p}
 \frac{\langle B_{ij}B_{mi}\rangle 
\langle B_{jl}B_{lm}\rangle }{\varepsilon_j-\varepsilon_m} - \nonumber \\
& & -\frac{\epsilon^4}{2}
 \sum_{i=1}^p 
\sum_{j=1}^p \sum_{n > p}\sum_{m > p} 
\frac{\langle B_{im}B_{jn}\rangle \langle B_{mi}B_{nj}\rangle }
{\varepsilon_j+\varepsilon_i-\varepsilon_m-\varepsilon_n}
\eea

This expansion is the continuous-time, Gaussian white noise equivalent of the  
expression obtained by Derrida\cite{Derrida87}.

\subsection{Parisi-Vulpiani}

Another model that can be revisited is the random matrix model
 introduced by
 Parisi and Vulpiani \cite{Parisi86} (see also Crisanti
 \cite{Crisanti93})
 in order to mimic some systems that show strong chaos.
A continuous-time version of this model, in the case of only one
 spatial degree
 of freedom ($N=2$) is a linearized  evolution of the form
\begin{equation}
V(t)=\left(\begin{matrix}0 &1 \cr \eta(t) & 0 \end{matrix}\right)
\label{priv}
\end{equation}
where $\eta$ is an Gaussian noise with a mean $r$ and a
 deviation
 $\sigma$.

Using this definition we can compute the average of 
\begin{equation}
\langle R^q(t) \rangle = \int P(\eta) d \eta \langle \psi^q|
{\cal {T}}\left[e^{\int_0^t H(\eta(t'))dt'}\right]|
\psi^{q\dag}\rangle 
\label{parvul1}
\end{equation}
where
\begin{equation}
 \psi^q=\prod^q_{{\gamma}=1}(a^{\gamma}_{q} b^{\gamma}_{q} + 
a^{\gamma}_{p}b^q_{p})
\end{equation}
and
\begin{equation}
H=\sum_{{\gamma}=1}^q(a_q^{{\gamma}\dag} a_p^{\gamma}
    + b_q^{{\gamma}\dag} b^{\gamma}_p )+\eta(t)
    \sum_{{\gamma}=1}^q(a_p^{{\gamma}\dag}
 a_q^{\gamma}
    + b_p^{{\gamma}\dag} b^{\gamma}_q )
\end{equation}
This Hamiltonian form of the exponential quantities associated with the 
Lyapunov exponents can be checked to be true  using small-time 
developments in the Suzuki-Trotter formula, just as in the previous
section.
 We can integrate the noise in (\ref{parvul1}) and obtain:
\begin{equation}
\langle R(t)^q\rangle = \langle \psi^q|
{\cal {T}}\left[e^{t H'}\right]|\psi^{q\dag}\rangle 
\label{parvul2}
\end{equation}
where the averaged Hamiltonian can be again checked in the discretized version:
\bea
H'&=&\sum_{{\gamma}=1}^q(a_q^{{\gamma}\dag} a_p^{\gamma}
    + b_q^{{\gamma}\dag} b^{\gamma}_p )+r\sum_{{\gamma}=1}^q
(a_p^{{\gamma}\dag} a_q^{\gamma}
    + b_p^{{\gamma}\dag} b^\gamma_q )+\nn 
&+&\frac{\sigma^2}{2} \left(\sum_{{\gamma}=1}^q(a_p^{{\gamma}\dag} a_q^{\gamma}
    + b_p^{{\gamma}\dag} b^{\gamma}_q )\right)^2
\label{pvav}
\eea

Let us first  concentrate on the zero average case $r=0$.
 We  make a redefinition of the 
fermions as:
\bea
a_q \rightarrow   t_1 a_q \;\;\; ; \;\;\;
 a_q^\dag \rightarrow  \frac{a_q^\dag}{t_1} 
\;\;\; &;& \;\;\;
a_p \rightarrow   t_2 a_p \;\;\; ; \;\;\;
 a_p^\dag \rightarrow \frac{a_p^\dag}{t_2} \nn  
\quad \frac{t_1}{t_2}&=&(\sigma^2)^{\frac{1}{3}}
\label{ren}
\eea
In terms of the new fermions, the dependence on $\sigma$ becomes explicit:
\begin{widetext}
\begin{equation}
H' \rightarrow  (\sigma^2)^{\frac{1}{3}} 
\left[ \sum_{{\gamma}=1}^q(a_q^{{\gamma}\dag} a_p^{\gamma}
    + b_q^{{\gamma}\dag} b^{\gamma}_p )+
\frac{1}{2} \left(\sum_{{\gamma}=1}^q(a_p^{{\gamma}\dag} a_q^{\gamma}
    + b_p^{{\gamma}\dag} b^{\gamma}_q )\right)^2 \right]  
\label{yy}
\end{equation}
\end{widetext}
We now have only  to diagonalize this time-independent Hamiltonian on 
subspaces with exactly two fermions of each replica.

We have done this numerically for values of $q$ up to 25 within the
$3^q$-dimensional basis generated by:
\begin{equation}
a_p^{\gamma \dag} b_p^{\gamma \dag} \;\;;\;\;
a_q^{\gamma \dag} b_q^{\gamma \dag} \;\;;\;\;
\frac{1}{\sqrt{2}}(a_p^{\gamma \dag} b_q^{\gamma \dag} + a_q^{\gamma \dag} b_p^{\gamma \dag})
\label{base}
\end{equation}
(symmetric under  Eq. (\ref{excha})).
For small $q$ it is easy to diagonalize $H$, for example for   $q=1$ we obtain 
$\lambda^2_1=(2 \sigma^2)^{\frac{1}{3}}$.

Let us note here that the  Hamiltonian (\ref{pvav}) is, for the
case $q=1$ the matrix 
\begin{equation}
\left(\begin{matrix}0&0&\sqrt{2}\cr\sigma^2&0&\sqrt{2}r\cr
 \sqrt{2}r&\sqrt{2}&0\end{matrix}\right)
\label{parvulmat}
\end{equation}
essentially the same used by
 Anteneodo and 
Vallejos \cite{Anteneodo02} to derive the same results.

\begin{figure}
\includegraphics[width=80mm]{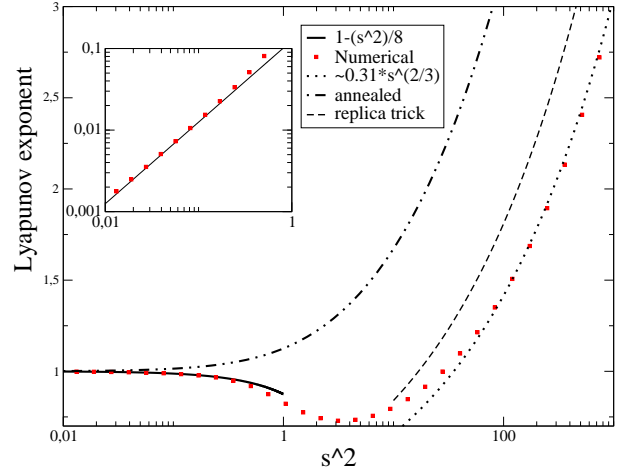}
\caption{\label{fig:parvul} 
 The greatest Lyapunov exponent versus the strength of the noise
 (s=$\sigma$)
 for a symplectic random matrix (the mean of the noise is 1). 
 At
 $\sigma >> 1$
 replica approximation and the power law fit are presented. The dashed
 line
 is the annealed Lyapunov exponent ($\frac{\lambda_1^2}{2}$).  For small $s$ , the Lyapunov exponent is given by $1-\frac{(s)^2}{8}$; the inset shows the check of this formula against the numerical data.}
\end{figure}

Let us now turn to the case  in which the noise has non-zero average.
We can transform  (\ref{pvav}) using
 transformations (\ref{ren})  with
\begin{equation}
\quad \frac{t_1}{t_2}=\sqrt{ r}
\label{renp}
\end{equation}
The Hamiltonian becomes $H'=\sqrt{r}H''$ with
\bea
H''&=& \sum_{{\gamma}=1}^q(a_q^{{\gamma}\dag} a_p^{\gamma} + b_q^{{\gamma}\dag} 
b^{\gamma}_p )+\sum_{{\gamma}=1}^q(a_p^{{\gamma}\dag} a_q^{\gamma} +
 b_p^{{\gamma}\dag} 
b^{\gamma}_q )+
 \nn&+&s^2 \left( \sum_{{\gamma}=1}^q(a_p^{{\gamma}\dag} 
a_q^{\gamma}+ b_p^{{\gamma}\dag} b^{\gamma}_q )\right)^2
\label{pvavr}
\eea
 where  and $s^2=
\frac{ \sigma^2}{r^{3/2}}$.
 There is a crossover
 (see Lima \cite{Lima90}) in the Lyapunov exponent dependence
 between the limits of small and large $s$ (see Fig. \ref{fig:parvul}).

Finally, let us  give an example on the role played by spatial structures.
  We consider a one dimensional system 
with a noise  $\eta$ (Eq. (\ref{priv})) 
 random variables  with correlations depending on the distance between sites:
\bea
\langle \eta_{ij}(t)\rangle &=&0 \nn \langle \eta_{ij}(t)\eta_{kl}(t')\rangle
 &=&\sigma^2(|i-j|)
(\delta_{ik}\delta_{jl}+\delta_{il}\delta_{jk})\delta(t-t') \nn
\eea
We shall compute the {\em annealed}
  Lyapunov exponent ($\lambda^2_1$). The Hamiltonian is
\bea
H=\sum_{l=1}^M(a_{q_l}^{\dag} a_{p_l}
    + b_{q_l}^{\dag} b_{p_l} )+ \sum_{l,m=1}^M\eta_{lm}(a_{p_l}^{\dag}
 a_{q_m}
    + b_{p_l}^{\dag} b_{q_m} )\nn
\eea
After averaging it becomes
\bea
H=\sum_{l=1}^M (a_{q_l}^{\dag} a_{p_l}+ b_{q_l}^{\dag} b_{p_l} )+
\sum_{l=1}^M \sigma^2(0) (a_{p_l}^{\dag} a_{q_l} + b_{p_l}^{\dag} b_{q_l})^2 \nn  +\sum_{k \neq l=1}^M 
\frac{\sigma^2(|l-k|)}{2}(a_{p_l}^{\dag} a_{q_k} + b_{p_l}^{\dag} 
b_{q_l} +a_{p_k}^{\dag} a_{q_l}  + b_{p_k}^{\dag} b_{q_l}^{\dag} )^2
\nn
\eea
The system becomes explicitly translational-invariant.
We have to diagonalize this Hamiltonian on the states 
\bea
\frac{1}{\sqrt{M}}\sum_{l=1}^M a_{q_l}^{\dag} b_{q_l}^{\dag} \nn
\frac{1}{\sqrt{M}}\sum_{l=1}^M a_{p_l}^{\dag} b_{p_l}^{\dag} \nn
\sum_{l=1}^M \frac{1}{\sqrt{2M}}(a_{q_l}^{\dag} b_{p_l}^{\dag}+
  a_{p_l}^{\dag} b_{q_l}^{\dag})
\eea
where $M$ is the total number of sites.
Within this subspace, $H$   is identical to the matrix (\ref{parvulmat}) 
with $r=0$ and we get:
\begin{equation}
\sigma^2=M (\sum_{i=1}\sigma^2(i)+2\sigma^2(0))=M\sum_{j=1}\langle \eta_{ij}\eta_{ij}\rangle 
\end{equation}
 again a result obtained before in Refs. \cite{Parisi86,Anteneodo02}.

 \section{Hamiltonian Mean Field system}
\label{HMF}
The Hamiltonian mean field  system we consider \cite{Antoni95}
is composed
  of $N$ 
 coupled rotators with a classical hamiltonian:
\begin{equation}
H=\sum_i\frac{p_i^2}{2}+\frac{J}{N}\sum_{i,j}[1-\cos(q_i-q_j)]
\end{equation}
This system has (in the canonical ensemble) a phase transition
 at $T_c=0.5 J$ between a ferromagnetic and a paramagnetic phase
 \cite{Antoni95,Latora99,Latora00}

The equations of motion are, if we allow for  noise and dissipation:
\bea
\frac{dq_i}{dt}&=&p_i\nn
\frac{dp_i}{dt}&=&-J M_x \sin q_i  +J M_y \cos q_i  - \gamma p_i +
 \sqrt{2 \gamma T} \eta_i\nn
\eea
where we have introduced:
\begin{equation}
M_x \equiv \frac{1}{N} \sum_i \cos q_i \;\;\; ; \;\;\;  
M_y \equiv \frac{1}{N} \sum_i \sin q_i
\end{equation}

The  matrix $V_{ij}$ governing the evolution operator in tangent space is:
\begin{equation}
V=\left(\begin{matrix}\mathbf{0}&\mathbf{1}\cr-E&-
\gamma\mathbf{1}\end{matrix}\right)
\end{equation}
with
\bea
 E_{ij}= -\frac{J}{N} \cos(q_i-q_j) \quad for \quad i\neq j \nn
E_{ii}=\frac{J}{N}\sum_i \cos(q_i-q_j)- \frac{J}{N}=\nn=J M_x\cos q_i +J M_y
\sin q_i  -\frac{J}{N} 
\eea
We can use now the function defined in (\ref{fon}) in order to compute the 
Lyapunov spectrum. 
The supersymmetry Hamiltonian becomes a sum of single particle terms:
\begin{eqnarray}
 H_S &=& \sum_i H_S^{(i)} \nonumber \\
  H_S^{(i)} &\equiv& \frac{\partial}{\partial p_i}
\left[J M_x 
\sin q_i - J M_y \cos q_i - \gamma p_i\right]\nonumber \\ 
& &+ T \gamma 
\frac{\partial^2}{\partial p_i^2}   -p_i \frac{\partial}{\partial q_i}
- J \nu_{q_i}^{l\dag} \nu^l_{p_i}\nonumber \\
& & + \gamma  \nu_{p_i}^{l \dag}
 \nu^l_{p_i}  + J (M_x \cos q_i + M_y \sin q_i)
\nu_{p_i}^{l \dag} \nu^l_{q_i} 
\nonumber \\ & & - J \nu^{l \dag}_{p_i} \cos q_i {\cal{C}}^l
-J \nu^{l \dag}_{p_i} \sin q_i {\cal{S}}^l
\label{single}
\end{eqnarray}
(here no summation over $i$ is implied).
For compactness we have defined $\nu^l$ for $l=1,2,3,4$:
\begin{eqnarray}
\nu_{q_i}^l &\equiv& (a_{q_i}, b_{q_i},\alpha_{q_i},\beta_{q_i}) 
\nonumber \\
\nu_{p_i}^l &\equiv& (a_{p_i}, b_{p_i},\alpha_{p_i},\beta_{p_i}) 
\end{eqnarray}
and the collective operators:
\begin{equation}
{\cal{C}}^l \equiv \frac{1}{N} \sum_i \nu_{q_i}^l \cos q_i \;\;\; ; \;\;\;  
{\cal{S}}^l  \equiv \frac{1}{N} \sum_i \nu_{q_i}^l  \sin q_i
\end{equation}
Not surprisingly, the operator $H_S$ is a mean-field (quantum) operator
itself, and 
we can solve the problem with any of the usual methods.
For example, introducing explicitly the collective 
variables using the functional delta-functions:
\begin{widetext}
\begin{eqnarray}
& &\int \; D[M_x] \; \delta(NM_x-\sum_i \cos q_i) = 
\int \; D[M_x]D[{\hat M}_x] \; e^
{-{\hat M}_x(NM_x-\sum_i \cos q_i)} \nonumber \\
& &\int \; D[M_y] \; \delta(NM_y-\sum_i \sin q_i) = 
\int \; D[M_y]D[{\hat M}_y] \; e^
{-{\hat M}_y(NM_y-\sum_i \sin q_i)} \nonumber \\
& &\int \; D[{\cal{C}}^l] \; 
\delta(N {\cal{C}}^l-\sum_i \nu_{q_i}^l \cos q_i) =
 \int \; D[{\cal{C}}^l] D[{\cal{{\hat C}}}^l] \; e^{-{\cal{{\hat C}}}^l
(N {\cal{C}}^l-\sum_i \nu_{q_i}^l \cos q_i)}\nonumber \\
& &\int \; D[{\cal{S}}^l] \; 
\delta(N {\cal{S}}^l-\sum_i \nu_{q_i}^l \sin q_i) =
 \int \; D[{\cal{S}}^l] D[{\cal{{\hat S}}}^l] \; e^{-{\cal{{\hat S}}}^l
(N {\cal{S}}^l-\sum_i \nu_{q_i}^l \sin q_i)}
\end{eqnarray}

we can write:
\begin{equation}
\langle  Z(\mu,{\bar\mu})\rangle =\int \; D[M_x]D[{\hat M}_x] \prod_l  D[{\cal{C}}^l] D[{\cal{{\hat C}}}^l] D[{\cal{S}}^l] D[{\cal{{\hat S}}}^l] \;e^{-N (\int dt \; 
({\hat M}_x M_x+
{\hat M}_y M_y+ {\cal{\hat C}}^l 
{\cal{C}}^l + {\cal{\hat S}}^l {\cal{S}}^l) - W)}
\label{zsaddle}
\end{equation}
\end{widetext}
where $W$ is the  action for a single pair of variables $(q_i,p_i)$:
 \begin{equation}
e^W=  \langle \phi_{L}^{(i)}| 
 \;{\cal T}\;  e^{-t (H_{eff}^{(i)}+\mu N_{ fer}^{(i)}+{\bar \mu} 
N_{bos}^{(i)})}(-1)^{N_{bos}^{(i)}}  |\phi_{R}^{(i)}\rangle 
\label{formulaas} 
\end{equation}

(no summation). 
The left and right eigenvectors are
\begin{eqnarray}
 |\phi_{R}^{(i)}
\rangle &=&\int d{ x_i}\;P({ x_i})\; e^{f^{(i)\dag}+  {\bar f}^{(i)\dag}}
|-\rangle \otimes 
|{ x_i}\rangle \nn
\langle \phi_{L}^{(i)}| &=& 
 \langle -|\otimes \langle -| e^{f^{(i)}+  {\bar f^{(i)} }} 
\end{eqnarray}
where for simplicity we have assumed that all pairs of variables $(q_i,p_i)$ 
have the same initial distribution, 
and hence left and right vectors are in product form.
The single particle effective Hamiltonian ($H^{(i)}_{eff}$) equals:
\begin{equation}
H^{(i)}_S - {\hat M}_x  \cos q_i - {\hat M}_y  \sin q_i - 
{\cal{{\hat C}}}^l  \nu_{q_i}^l \cos q_i
-{\cal{{\hat S}}}^l  \nu_{q_i}^l \sin q_i
\label{Heff}
\end{equation}
which has  in principle a time dependence through the collective variables.
We wish to evaluate $<Z(\mu,\mu')>$ by  saddle-point method in the
thermodynamical limit. 
The saddle point equations for the ordinary collective variables read:
\begin{eqnarray}
\hat {M_x}&=&\frac{\partial W}{\partial M_x} \quad 
 M_x=\frac{\partial W}{\partial \hat{M_x}} \nonumber \\
\hat {M_y}&=&\frac{\partial W}{\partial M_y} \quad 
 M_y=\frac{\partial W}{\partial \hat{M_y}}
\label{8eqs}
\end{eqnarray}
We also have saddle point equations  for the sixteen fermionic and 
bosonic collective variables $ {\cal{{\hat C}}}^l
,{\cal{{ C}}}^l,
{\cal{{\hat S}}}^l,{\cal{{ S}}}^l$.

  We have  20 equations of type (\ref{8eqs}) which we can solve
 assuming  that the system
 is at $t=0$  already in thermodynamic equilibrium -- 
the vector to the right in (\ref{formulaas})
is Gibbs-distributed. This implies that the collective variables
 may be constant in time, an assumption we verify later.
First of all, it is easy to see that the 
saddle-point values of the 
$ {\cal{{\hat C}}}^l
,{\cal{{ C}}}^l,
{\cal{{\hat S}}}^l$ and ${\cal{{ S}}}^l$ vanish (Appendix \ref{hmfmf}).
 Next, one can see  that if the 
endpoint of the trajectories is left free 
(i.e. we are not conditioning to a specific arrival point),
then causality implies that $\hat M_x=\hat M_y=0$ (Appendix \ref{hmfmf}).
Under these assumptions, 
the only variables with non-zero saddle-point values are:
\begin{equation}
 M_x= \langle \cos q_i \rangle_M \;\;\; ; \;\;\;
 M_y=\langle \sin q_i \rangle_M
\label{8eqs8}
\end{equation}
where the average $\langle \bullet \rangle_M$ is taken
with the single particle dynamics:
\bea
\frac{dq}{dt}&=&p\nn
\frac{dp}{dt}&=&-JM \sin q - \gamma p + \sqrt{2 \gamma T} \eta
\label{spp}
\eea
Indeed, this yields the equilibrium value of $M$, as solution of the equation: \begin{equation}
M=\frac{I_1(y)}{I_0(y)} \quad y=\frac{MJ}{T}
\end{equation}
(where $I$ are the Bessel  functions, see \cite{Antoni95}). 
What we have shown is that the Lyapunov density function 
 of the system is, to leading order in $N$
the sum of Lyapunov densities for a single particle
 system moving according to Eqn. (\ref{spp}).

In the  limit of zero coupling to the bath $\gamma=0$ the 
exponents for a single  particle 
are zero,  as a consequence of
      conservation of  energy and  the pairing rule. Note that the fact
that the energy of each particle is conserved separately is
an artifact of the large $N$-limit: there is a coupling between 
the fluctuations at the following order in $1/N$.
 In conclusion, in the thermodynamical
 limit the function $G(\mu)$ is a step of height $\sim 2N$ at zero.

A  vanishing largest Lyapunov exponent $\lambda_i$
   has already been
 obtained both numerically 
(\cite{Latora99, Latora00}) and analytically  (\cite{Anteneodo02, Firpo98}) 
in the paramagnetic phase, but not  in the ferromagnetic phase.
 This is not in contradiction with our results:
 according to our calculation
one can still have a vanishing fraction 
of non-zero Lyapunov exponents,  that do not contribute to the density 
function in the large-$N$ limit.
If we wish
 to calculate the largest exponent $\lambda_1$ we should take
 our calculation
to the next order.
\begin{figure}
\includegraphics[width=8.5cm]{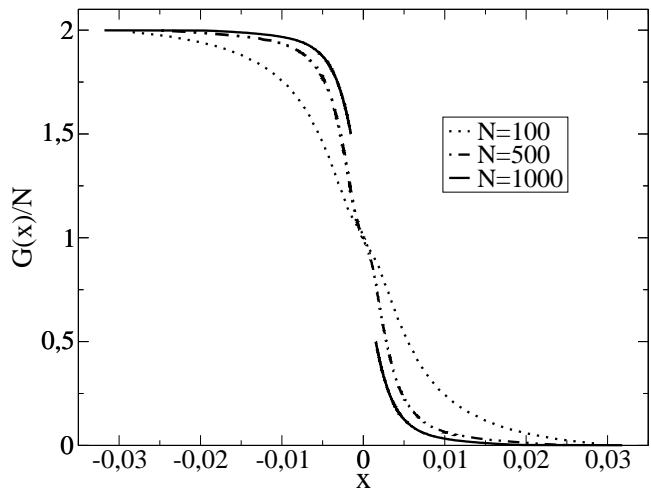}
\caption{\label{fig:step} 
{The microcanonical Lyapunov spectra of the HMF system with 
$N$ particles at the energy $E=0.3$ presented as $\frac{G(x)}{N}$
(see equation (\protect\ref{cdef})) for different numbers of particles ($N$). Only the 
first half of the spectra is computed, the other half being obtained by the 
pairing rule. }}
\end{figure}

Figure \ref{fig:step} shows a numerical calculation of the spectrum of
 Lyapunov exponents for various values of the number of particles $N$. 
The simulations are 
carried out in the microcanonical ensemble at an energy where the greatest Lyapunov exponent 
does not show an important variation with $N$ ($E=0.3$, see \cite{Latora99}).
Even though the computer time requirements do not allow us to get close to 
the thermodynamical limit, these numerical results show a reasonably good 
agreement with the hypothesis of a step spectrum.

\begin{figure}
\includegraphics[width=8.5cm]{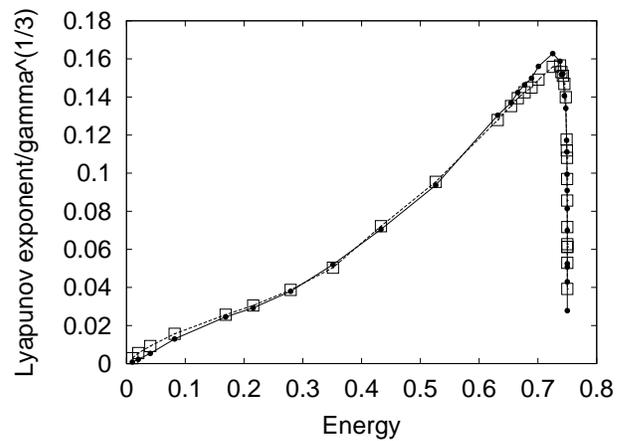}
\caption{\label{fig:ruflyap} {Lyapunov exponent for one particle
 with Kramers dynamics
 (\protect\ref{spp}) versus the energy for $\gamma=.01$ (dots) and $\gamma=.001$ (squares) respectively. $E=0.75$ is the energy at the transition point (for $J=1$) 
and the Lyapunov exponent is automatically zero above this point.
The exponents are scaled with $\gamma^{\frac{1}{3}}$.}}
\end{figure}

At finite $\gamma$ the Lyapunov distribution function is no longer
a step, even in the thermodynamic limit. 
We can obtain it very easily by solving  the one-particle 
system (\ref{spp}) numerically.
The results depend strongly of $\gamma$; there is a scaling law
 in $\gamma^{1/3}$ below the critical temperature and an identical
 zero $\lambda_1$  in the paramagnetic phase (see Fig. \ref{fig:ruflyap}).

Surprisingly, after rescaling the largest Lyapunov  exponent
 behaves very much like the one  computed by Firpo (\cite{Firpo98})
 in the canonical ensemble. There are also qualitative and quantitative 
resemblances with the  one obtained in micro-canonical simulations
 of a large (but obviously finite) number of particles (see \cite{Latora99, Latora00}). 
We can conclude that, at finite $\gamma$, $G(\mu)$ should have two sigmoids of 
height  $\sim N$ located symmetrically around zero.

%\vspace{.5cm}
\section{Conclusions}
\label{conclusions}
%\vspace{.5cm}

 Writing the  Lyapunov exponents as a statistical mechanic object 
provides a different perspective of the problem, allowing 
to transpose 
much of the knowledge and intuition developed in that wider context.

Perhaps the most clear example
are the questions related to the thermodynamic
limit. We have seen that the existence of a limit Lyapunov density
function is the kind of extensivity property that most theoretical physicists
would accept without proof --- at least for the times that do not diverge with
the system size.

As to the different approximation schemes, a case in point is 
the continuous product of random matrices Section \ref{RM}.
We have seen there that the problem is mapped
into  a system of interacting fermions.
 We have treated it as a
perturbative expansion of weak interaction
reobtaining the results of 
 Ref. \cite{Derrida87}. However, once  in the language of interacting
particles, one immediately thinks of other, more global approximations;
we have already mentioned Hartree-Fock, which can easily be implemented both
in the pure fermion or in the supersymmetric formalism.
Similarly, the standard mean-field treatment of the model
of section \ref{HMF} can be extended  for the calculation of the exponents.
This also suggests that it may be interesting in general to construct
a local mean-field approximation for problems with space: this could
give a simple analytic handle on the spatial structures involved
with each exponent. 

A question into which we have not looked  in detail is  intermittency.
The generating function of generalized  Lyapunov exponents $\lambda^{2q}_i$
involves a distribution
function in space that depends on $q$.
 Physically,
this arises because we are conditioning the probability of a trajectory to
having an unusual value of the exponents. Hence, studying the
  lowest eigenvector of $H_q$   gives us information on the
spatial structures responsible for  intermittent behavior.

Yet another interesting question is to look at systems at or near criticality,
and borrow methods and ideas from 
the rich theory of critical phenomena to infer results on the behavior
of the exponents there.

\vspace*{1cm}
{\bf \Large Acknowledgment}
\vspace*{.5cm}

We wish to thank  Stefano Ruffo for clarification and encouragement at all stages of the 
preparation of this work.

\appendix
\section{}
\label{A}
In this appendix we show how the use of fermions transform the quantity $R$ defined in (\ref{nv}) can be written as a certain matrix element of the hamiltonian (\ref{ham22}) (equation \ref{nvf})).

Let $a^\dag_i$ be a set of fermion operators 
\begin{equation}
[a_i,a^\dag_j]_+=\delta_{ij}
\end{equation}
We can encode (\ref{nv})  by writing:

\begin{equation}
O(t)= a^\dag_k y_k(t)
\label{ot}
\end{equation}

We put:
\begin{equation}
\dot{O}(t) = [ V_{kl} a^\dag_k a_l \; , O(t) ]_-
\label{tres}
\end{equation}

It is easy to check that the $y_i$ then satisfy (\ref{te}). The solution
to (\ref{tres}) is:
\begin{equation}
O(t)= \left( {\cal {T}} e^{t V_{kl} a^\dag_k a_l}\right)
 \; O(0) \; 
\left({\cal {T}} e^{t V_{k'l'} a^\dag_{k'} a_{l'}} 
\right)^{-1}
\label{cuatro}
\end{equation}
which implies:
\begin{equation}
 \left({\cal {T}} e^{ t V_{kl} a^\dag_k a_l}\right) \;a^\dag_i \; 
\left({\cal {T}} e^{  t V_{k'l'} a^\dag_{k'}
 a_{l'}}\right)^{-1}   = U_{ji} a^\dag_j
\label{cinco}
\end{equation}

Let us now add a second set of fermions $b^\dag_k$, and construct the operator:
\begin{equation}
f \equiv a^\dag_k b^\dag_k
\label{seis}
\end{equation}
so we can write $ U_{ji} U_{ri} a^\dag_j b^\dag_r $ as:
\begin{equation}
 \left( {\cal {T}} e^{ t V_{kl} (a^\dag_k a_l  + b^\dag_k b_l    )}\right)
 \; f \; 
\left({\cal {T}} e^{t V_{k'l'} (a^\dag_{k'} a_{l'} + b^\dag_{k'} b_{l'})} 
\right)^{-1} 
\label{siete}
\end{equation}
(note that the same constructions can be made with bosons instead of fermions.)
Denoting $|-\rangle $ the fermion vacuum, Eq. (\ref{siete}) implies:

\begin{widetext}
\bea
\langle -|a_{i_1} ...a_{i_p}  b_{i_1} ... b_{i_p}|{\cal {T}}\left[e^{V_{ij}(a_i^\dag a_j
    + b_i^\dag b_j )}\right]|
a^\dag_{l_1}...a^\dag_{l_p} b^\dag_{l_1}...b^\dag_{l_p}|-\rangle =\nn
U_{k_1l_1}...U_{k_p l_p}U_{j_1l_1}...U_{j_pl_p}\langle -|a_{i_1} ...a_{i_p}
a^\dag_{k_1}...a^\dag_{k_p}|-\rangle \langle -|b_{i_1} ... b_{i_p}
 b^\dag_{j_1}...b^\dag_{j_p}|-\rangle 
\label{desf}
\eea
\end{widetext}

Summing over $i_1,...,i_p$ the first part transforms into:
\beq
U_{k_1l_1}U_{j_1l_1}...U_{k_p l_p}U_{j_pl_p}=(UU^\dag)_{k_1j_1}...(UU^\dag)_{k_p j_p}
\label{aa}
\eeq

Denoting with $\sigma((j_1,...,j_p)\rightarrow (k_1,...,k_p))$ the signature 
of the permutation between the two sets of indices the second (fermionic) part 
of (\ref{desf}) equals
\begin{equation}
\sigma((i_1,...,i_p)\rightarrow (k_1,...,k_p))\sigma((j_1,...,j_p)\rightarrow (i_1,...,i_p))
\end{equation}

Using (\ref{aa}) (\ref{desf}) and the properties of the signature it follows that  (\ref{desf}) equals

\begin{equation}
(p!)\sum_{j_1,k_1,....j_p,k_p}A_{j_1k_1}...A_{j_n k_p}
\sigma((j_1,...,j_p)\rightarrow (k_1,...,k_p))
\label{ij}
\end{equation}
(remember that $A=U^\dag U$).
After summing on  $k_1,...,k_p$ the expression (\ref{ij}) is, by definition:
\begin{equation}
(p!)^2 det\left[(y_{i_1}^\dag,.....,y_{i_p}^\dag)^\dag\left[
U^\dag U \right](y_{i_1},.....,y_{i_p}) \right]
\end{equation}

\section{}
\label{bosons}
In this appendix we derive formula (\ref{fon}).
Denoting $|-\rangle $ the fermion vacuum, Eq. (\ref{siete}) implies:
\begin{equation}
 \left( {\cal {T}} e^{t V_{kl} (a^\dag_k a_l  + b^\dag_k b_l    )}\right)
\; e^{f} |-\rangle = e^{ A_{jr}  a^\dag_j b^\dag_r}|-\rangle 
\label{ocho}
\end{equation}
and:
\begin{eqnarray}
 \langle -| e^{ f^\dag} \left( {\cal {T}} e^{t V_{kl} (a^\dag_k a_l  +
 b^\dag_k b_l  )  -\mu N_{fer} t}\right)
\; e^{f} |-\rangle = \nn = 
 \langle -|e^{ b_i a_i }e^{ -\mu N_{fer}} e^{\gamma A_{jr}  a^\dag_j b^\dag_r}|-\rangle  
=  det [ 1 + e^{-\mu t}A]
\label{nueve}
\end{eqnarray}
The last equality is most easily checked by performing a rotation of
the fermions to diagonalize $A$. Then we have:
\bea
& \langle -|e^{ b_i a_i }e^{ -\mu N_{fer} t} e^{ A_{jr}  a^\dag_j b^\dag_r}|-\rangle&=\nn
&\langle -|e^{ b^*_i a^*_i e^{-\mu t}}e^{ {\cal{A}}_{j}  a^{*\dag}_j
 b^{*\dag}_j}|-\rangle&=\nn &\prod_k \langle -|e^{ b^*_k a^*_k e^{-\mu t}} 
e^{ {\cal{A}}_{k}  a^{*\dag}_k b^{*\dag}_k}|-\rangle&
\label{xx}
\eea
where ${\cal{A}}$ are the eigenvalues of $A$ (no summation on k); in the last expression
 there is 
no sum on $i$ in the exponents. Each factor can be developed into
 $1+e^{-\mu t}{\cal{A}}_i$, so the whole expression will be:
\begin{equation}
\prod_i(1+e^{-\mu t}{\cal{A}}_i)= det [ 1 + e^{-\mu t} A]
\label{xxxx}
\end{equation}

Consider now the same steps, but now replacing the fermions $a^\dag_k$ and
$b^\dag_k$ by bosons $\alpha^\dag_k$ and $\beta^\dag_k$ as in (\ref{mann});
then we obtain the analogue of (\ref{nueve}):
\begin{eqnarray}
 \langle -| e^{ {\bar f}^\dag} 
\left( {\cal {T}} e^{t V_{kl} (\alpha^\dag_k \alpha_l  + \beta^\dag_k \beta_l 
   )-\mu N_{bos} t}\right)
\; e^{{\bar f}} |-\rangle = \nn = 
 \langle -|e^{ \beta_i \alpha_i e^{- \mu t}} e^{ A_{jr} 
 \alpha^\dag_j \beta^\dag_r}|-\rangle  = det [ 1 - e^{- \mu t} A]^{-1}
\label{nuevep}
\end{eqnarray}
which can again be proved by performing a rotation of the bosons to
diagonalize $A$. 
After the same step as in (\ref{xx}) the terms that survive are those with 
an equal number of the creation and destruction operators and we can use
\begin{equation}
\langle -|(\alpha^n\beta^{\dag n})|-\rangle =n!
\end{equation}
to obtain the analogue of (\ref{xxxx}):
\begin{equation}
\prod_i(1+   e^{-\mu t}{\cal{A}}_i+ 
e^{-2 \mu t}{\cal{A}}_i^2+...)=\prod_i 
\frac{1}{ 1 - e^{-\mu t}{\cal{A}}_i }
\label{bip}
\end{equation}
One can change the sign in (\ref{bip}) we can introduce some ``negative norm'' 
bosonic states, or as:
\bea
\frac{1}{det[1+A]}= \langle -| e^{ {\bar f}^\dag} 
\left( {\cal {T}} e^{t V_{kl} (\alpha^\dag_k \alpha_l  + \beta^\dag_k \beta_l  
  )}\right) e^{- {\bar f}} |-\rangle \nn=
 \langle -| e^{ {\bar f}^\dag} 
\left( {\cal {T}} e^{t V_{kl} (\alpha^\dag_k \alpha_l  + \beta^\dag_k \beta_l 
   )}\right) (-1)^{N_{bos}}e^{{\bar f}} |-\rangle 
\eea

\section{}
% BOREL SUMMATION
\label{borsum}
In this appendix we show the ladder structure of the functions $G$ defined in
 (\ref{fon2}) and (\ref{fon2b}). The common ingredient after 
developing the exponentials and matching the terms allowed by the conservation of bosons and fermions is:

\begin{widetext}
\begin{equation}
\langle -|(f^\dag+  {\bar f}^\dag)^n
\left( {\cal {T}}(-1)^{N_{bos} }
e^{t  V_{kl} (a^\dag_k a_l+b^\dag_k b_l)+tV_{kl}(\alpha^\dag_k 
\alpha_l  + \beta^\dag_k \beta_l ) - \mu t(N_{bos}+N_{fer})}\right)
N_{fer}(f+  {\bar f})^n|-\rangle 
\label{bon2}
\end{equation}
 We can commute the ``evolution'' operator with $N_{fer}$ and  apply 
it on the right ket; the action of the operators number of bosons and fermions 
$N_{bos}+N_{fer}$ (on the left ket) can be easily computed so the expression becomes:
\begin{equation}
e^{-n \mu t} \langle -|\frac{(f^\dag- {\bar
    f}^\dag)^n}{n!}N_F \frac{({\cal{A}}_ia^{\dag}_ib^{\dag}_i+ {\cal{A}}_i 
\alpha^{\dag}_i\beta^{\dag}_i )^n}{n!}|-\rangle 
\end{equation}
The expansion of the powers in terms of individual creators an destructors:
\begin{equation}
e^{-n\mu t}(n!)^2\sum_{k_1+...+k_N=n}\langle -|\prod_j
\frac{(b_ja_j-\beta_j\alpha_j)^{k_j}}{k_j!}N_{fer}
\prod_j({\cal{A}}_j)^{k_j}\frac{(a_j^{\dag}b_j^{\dag}+\alpha_j^{\dag}
\beta_j^{\dag})^{k_j}}{k_j!}|-\rangle 
\label{bon1}
\end{equation}
Here there is no sum on repeated index inside the  parenthesis. 
Expanding  further $N_{fer}$ in terms of fermions operators this term
 can be written as:
\bea
 e^{-n\mu t} (n!)^2 \sum_l \sum_{\sum k_i=n}
\langle -|
\frac{(b_la_l-\beta_l\alpha_l)^{k_l}}{k_l!}
\frac{a^{\dag}_la_l+b^{\dag}_lb_l}{2}
{\cal{A}}^{k_l} 
\frac{(a_l^{\dag}b_l^{\dag}+\alpha_l^{\dag}\beta_l^{\dag})^{k_l}}{k_l!}
\prod_{j\neq l}\frac{(b_ja_j-\beta_j\alpha_j)^{k_j}}{k_j!}
{\cal{A}}_j^{k_j}
\frac{(a_j^{\dag}b_j^{\dag}+\alpha_j^{\dag}\beta_j^{\dag})^{k_j}}{k_j!}
|-\rangle 
\eea
But for each $l$, at $j\neq l$ and $k_j > 0$ a factor of the product above 
will be:
\bea
(-{\cal{A}}_j)^{k_j}\left(
\frac{\langle -|\beta_j^{k_j}\alpha_j^{k_j}\alpha_j^{\dag k_j}\beta_j^{\dag
    k_j}|-\rangle }{k_j!^2}-\frac{k_j^2\langle -|\beta_j^{k_j-1}\alpha_j^{k_j-1}
\alpha_j^{\dag k_j-1}\beta_j^{\dag k_j-1}|-\rangle 
\langle -|b_ja_ja_j^{\dag}b_j^{\dag}|-\rangle }{k_j!^2})\right)
\eea

and we can easily see that this equals zero.
 So, inside each term of the sum on l, only that with
$k_l=n$ and the others $k_j=0$ will survive.

The sum (\ref{bon1}) simplifies to:
\begin{equation}
 e^{-n\mu t}(n!)^2 \sum_l 
\frac{{\cal{A}}_l^{n}(-1)^{n-1}n^2\langle -|\beta_l^{n-1}\alpha_l^{n-1}\alpha_l^{\dag
    n-1}\beta_l^{\dag
    n-1}|-\rangle \langle -|b_la_l\frac{a^{\dag}_la_l+b^{\dag}_lb_l}{2}
a_l^{\dag}b_l^{\dag}|-\rangle }{n!^2}= e^{-n\mu t} \sum_l  {\cal{A}}_l^{n}(-1)^{n-1}
\label{borsum1}
\end{equation}
\end{widetext}
Now we can  reconstruct the sum on n for the different cases; 
the Borel construction gives uses the series
\begin{equation}
-\sum_{j=1}^{N} \sum_{n=1}^\infty\frac{(-e^{-\mu t} 
{\cal{A}}_j)^n}{n!}=\sum_{j=1}^{N}[1-exp(-e^{-\mu t}{\cal{A}}_j)]
\label{c7}
\end{equation}
The series from (\ref{c7}) are convergent at each time and for each 
trajectory (in a well behaving, smooth potential); one can formally sum the 
general terms (\ref{borsum1}) without the $\frac{1}{n!}$ and obtain:
\begin{equation}
-\sum_{j=1}^{N} \sum_{n=1}^\infty(-e^{-\mu t} {\cal{A}}_j)^n=
\sum_{j=1}^{N}\frac{e^{-\mu t} {\cal{A}}_j}{1+e^{-\mu t} {\cal{A}}_j}
\label{c9}
\end{equation}
Using the exponential form of ${\cal A}_j$, in the limit 
$t \rightarrow \infty$, both (\ref{c7}) and (\ref{c9}) will give
\beq
G(\mu)=G^B(\mu)=\sum_{j=1}^{N}\theta(\lambda-\mu)
\label{gbg}
\eeq

\section{}

In this appendix we derive the fourth term in $\epsilon$ of
 the weak disorder expansion treated in section (\ref{derrida}).
All the eigenstates of $H_o$ can be obtained by destroying
 and creating 
fermions in $\psi_o$. 
Due to the fact that the perturbation contains two creators
 and two destructors the states connected  by the perturbation
 with $\psi_o$ are only those 
that differ from it by one or two fermions. 
The ``replica composition'' of the states connected by the
 perturbation must be the same (two different replicas are not connected).

One set of states is formed by destroying one fermion from the
 first $p$ and creating one (from the last $N-p$); both fermions
 must be in the same replica family: 
\bea
\psi_{b\gamma jm}=b^{\gamma\dag}_j b^\gamma_m \psi_0 \nn
\psi_{a\gamma jm}=a^{\gamma\dag}_j a^\gamma_m \psi_0 \nn
 j \le p \quad m > p
\eea

These eigenstates will contribute to the Lyapunov exponent by a second order
 perturbation term:
\bea
\sum_{\gamma=1}^q \sum_{j=1}^p \sum_{m> p} 
\frac{\langle \psi_{b\gamma jm}|V|\psi_0\rangle \langle \psi_0|V|\psi_{b\gamma jm}\rangle 
 }{\varepsilon_j-\varepsilon_m}
 + \\  \sum_{\gamma=1}^q \sum_{j=1}^p \sum_{m> p}
 \frac{\langle \psi_{a\gamma jm}|V|\psi_0\rangle \langle \psi_0|V|
\psi_{a\gamma jm}\rangle  }{\varepsilon_j-\varepsilon_m}
\eea
The linear part in q of this term is:
\begin{equation}
2q \sum_{i=1}^p \sum_{l=1}^p \sum_{j=1}^p\sum_{m> p}
 \frac{\langle B_{ij}B_{mi}\rangle \langle B_{jl}B_{lm}\rangle }{\varepsilon_j-\varepsilon_m}
\end{equation}

The states that have two different fermions are of three types:
\bea
\psi_{b\gamma in;b \gamma' jm}=b^{\gamma\dag}_i b^{\gamma'\dag}_j 
b^{\gamma'}_m b^\gamma_n \psi_0 \nn
\psi_{a\gamma in;b \gamma' jm}=a^{\gamma\dag}_i b^{\gamma'\dag}_j
 b^{\gamma'}_m a^\gamma_n \psi_0 \nn
\psi_{a\gamma in;a \gamma' jm}=a^{\gamma\dag}_i a^{\gamma'\dag}_j 
a^{\gamma'}_m a^\gamma_n \psi_0 \nn
i \le p \quad j \le p \quad m >  p \quad n >  p 
\eea
and each of them will contribute with terms of the form
\begin{equation}
\sum_{\gamma=1}^q \sum_{j=1}^p \sum_{m> p} 
\sum_{\gamma'=1}^q \sum_{i=1}^p \sum_{n> p} 
\frac{\langle \psi|V|\psi_0\rangle \langle \psi_0|V|\psi\rangle  }{\varepsilon_j+
\varepsilon_i-\varepsilon_m-\varepsilon_n}
\end{equation}

The linear term in q will be:
\begin{equation}
q \sum_{i=1}^p \sum_{j=1}^p \sum_{n> p}\sum_{m> p} 
\frac{\langle B_{im}B_{jn}\rangle \langle B_{mi}B_{nj}\rangle }{\varepsilon_j+\varepsilon_i-\varepsilon_m-\varepsilon_n}
\end{equation}

\section{}
\label{hmfmf}
In this appendix we study the consistency
 equations for the collective variables in
 the HMF model. 
As we can see in from the equation (\ref{liapu1}) we need quantities at 
$\mu =\bar \mu$; the derivation implied by the definition of the $G$ function 
will be carried out on $<Z(\mu,\bar \mu)>$ after the saddle point evaluation.
So, we must take the derivative of the expression (\ref{zsaddle}). 
In this point, the derivative reads:
\begin{widetext}
\beq
 \left.\frac{\partial Z^S(\mu,{\bar\mu})>}{\partial \mu}\right|_{{\bar\mu}=
\mu}=\frac{1}{Z^S}\left[ \left.\frac{\partial W(\mu,{\bar\mu})}{\partial
 \mu}\right|_{{\bar\mu}=\mu}+\sum_k\left.\frac{\partial Z^S(\mu,{\bar\mu})>}{\partial {\cal{X}}_k}\frac{\partial {\cal{X}}_k}{\partial \mu}\right|_{{\bar\mu=\mu}}
\right]
\label{muemup}
\eeq
\end{widetext}
where we take  ${\cal{X}}_k$, for $k=1,...,20$, to be the vector of the
 collective variables, and $Z^S$ is the value of $<Z(\mu,\bar \mu)>$ at the saddle (\ref{zsaddle}).
 We see that the second part of the rhs of this equation
 is zero because of the consistency equations (\ref{8eqs}); in conclusion we 
are interested in 
 the values of the collective variables at $\mu=\bar \mu$ and 
only the direct dependence on the two variables of $W$ will be interesting 
for us. 
With this assumption we write the consistency equation for the boson and fermion variables as:
 \begin{equation}
{\cal{S}}^l= \frac{ \langle \phi_{L}^{(i)}| 
 \;{\cal T}\;  e^{-t (H_{eff}^{(i)}+\mu {\bar N})} \nu_{q_i}^l  \sin q_i (-1)^{N_{bos}^{(i)}} |\phi_{R}^{(i)}\rangle }{e^{W}}
\label{cons1} 
\end{equation}

 \begin{equation}
{\cal{\hat S}}^l= \frac{ \langle \phi_{L}^{(i)}| 
 \;{\cal T}\;  e^{-t (H_{eff}^{(i)}+\mu{\bar N} )}  \nu_{p_i}^{l\dag}  \sin q_i(-1)^{N_{bos}^{(i)}}  |\phi_{R}^{(i)}\rangle }{e^{W}}
\label{cons2} 
\end{equation}

 \begin{equation}
{\cal{ C}}^l= \frac{ \langle \phi_{L}^{(i)}| 
 \;{\cal T}\;  e^{-t (H_{eff}^{(i)}+\mu {\bar N})} \nu_{q_i}^l  \cos q_i(-1)^{N_{bos}^{(i)}}  |\phi_{R}^{(i)}\rangle }{e^{W}}
\label{cons3} 
\end{equation}
 \begin{equation}
{\cal{\hat C}}^l= \frac{ \langle \phi_{L}^{(i)}| 
 \;{\cal T}\;  e^{-t (H_{eff}^{(i)}+\mu {\bar N})} \nu_{p_i}^{l\dag}  \cos q_i (-1)^{N_{bos}^{(i)}} |\phi_{R}^{(i)}\rangle }{e^{W}}
\label{cons4} 
\end{equation}
All those equations contain expectation values  of {\em single}
 boson and fermion operators which necessarily vanish.
 This is normal, as we could expect 
from the beginning that those variables integrate out, contributing only 
with a non-exponential prefactor. This prefactor (as expected from the 
supersymmetry considerations ) will be one in the limit  $\mu=\bar \mu$ but in 
the process of derivation will give terms of order $O(\frac{1}{N})$.
 
There are four equations left can be divided in two parts. The first parts 
contains the  variables $\hat M_x$ and  $\hat M_y:$
\begin{widetext}
 \begin{equation}
\hat M_x= J\frac{ \langle \phi_{L}^{(i)}| 
 \;{\cal T}\;  e^{-t (H_{eff}^{(i)}+\mu {\bar N})}(\frac{\partial}{\partial p_i}
\sin q_i + \cos q_i 
\nu_{p_i}^{l \dag} \nu^l_{q_i} )
(-1)^{N_{bos}^{(i)}}  |\phi_{R}^{(i)}\rangle }{e^{W}}
\label{cons5} 
\end{equation}
 \begin{equation}
\hat M_x =-J \frac{ \langle \phi_{L}^{(i)}| 
 \;{\cal T}\;  e^{-t (H_{eff}^{(i)}+\mu {\bar N})}(\frac{\partial}{\partial p_i}
\sin q_i - \sin q_i 
\nu_{p_i}^{l \dag} \nu^l_{q_i} )
(-1)^{N_{bos}^{(i)}}  |\phi_{R}^{(i)}\rangle }{e^{W}}
\label{cons6} 
\end{equation}
The first term of the rhs of these equations is zero 
(as demanded by the causality) and the second is also zero, due to the 
mismatch in the fermion-boson operators.

Finally, the last two equations are the only nontrivial ones:
 \begin{equation}
 M_x= \frac{ \langle \phi_{L}^{(i)}| 
 \;{\cal T}\;  e^{-t (H_{eff}^{(i)}+\mu {\bar N})}\cos(q_i)(-1)^{N_{bos}^{(i)}} 
 |\phi_{R}^{(i)}\rangle }{e^{W}}
\label{cons7} 
\end{equation}
 \begin{equation}
 M_y = \frac{ \langle \phi_{L}^{(i)}| 
 \;{\cal T}\;  e^{-t (H_{eff}^{(i)}+\mu {\bar N})}\sin(q_i)(-1)^{N_{bos}^{(i)}} 
 |\phi_{R}^{(i)}\rangle }{e^{W}}
\label{cons8} 
\end{equation}
Now the hamiltonian $H_{eff}^{(i)}$ is much simpler:
 \begin{equation}
H_{eff}^{(i)}=\frac{\partial}{\partial p_i}
\left[J M
\sin q_i - \gamma p_i\right]+ T \gamma 
\frac{\partial^2}{\partial p_i^2}   -p_i \frac{\partial}{\partial q_i}
- J \nu_{q_i}^{l\dag} \nu^l_{p_i}+ \gamma  \nu_{p_i}^{l \dag}
 \nu^l_{p_i}  + J M \cos q_i 
\nu_{p_i}^{l \dag} \nu^l_{q_i} 
\label{heffs}
\end{equation}
\end{widetext}
where we used also the rotational symmetry (which define a saddle manifold) on the space of $M_x$ and $M_y$ which allow us to fix $M_y=0$ and $M_x=M$; looking back at (\ref{zsaddle}) and (\ref{muemup}) we can conclude that 
\beq
G(\mu)=N\lim_{t \rightarrow \infty} \frac{1}{t e^{-W}} \left.\frac{\partial W(\mu,{\bar\mu})}{\partial
 \mu}\right|_{{\bar\mu}=\mu}
\eeq
Formula (\ref{formulaas}) combined with the simple expression of 
$H^{(i)}_{eff}$ (\ref{heffs}) allow us to infer that 
\beq
G(\mu)=N G^{(i)}(\mu)
\eeq
where $G^{(i)}(\mu)$ characterize the Lyapunov spectrum of a single particle with the dynamics (\ref{spp}); this spectrum contains in fact only two exponents.

\bibliographystyle{apsrev}
\bibliography{biblio}

\end{document}